\documentclass[showpacs,aps,floatfix,pra,reprint,superscriptaddress]{revtex4-1}
\usepackage{enumitem}
\usepackage{amssymb}
\usepackage{graphicx}
\usepackage{verbatim}
\usepackage{amsfonts} 
\usepackage{amsmath}  
\usepackage{mathtools}
\usepackage{soul,xcolor,colortbl}
\usepackage{hyperref} \hypersetup{colorlinks=true,citecolor=blue,linkcolor=blue,urlcolor=blue} 
\usepackage{tabularx}
\usepackage{bm}    
\usepackage{array}
\usepackage{color} %

\usepackage{soul}
\usepackage[utf8]{inputenc} 
\DeclareUnicodeCharacter{2010}{-}
\newcolumntype{L}[1]{>{\raggedright\let\newline\\\arraybackslash\hspace{0pt}}m{#1}}
\newcolumntype{C}[1]{>{\centering\let\newline\\\arraybackslash\hspace{0pt}}m{#1}}
\newcolumntype{R}[1]{>{\raggedleft\let\newline\\\arraybackslash\hspace{0pt}}m{#1}}

\usepackage{microtype,booktabs}

\usepackage[utf8]{inputenc}
\usepackage[T1]{fontenc}
\usepackage{textcomp}
\usepackage{comment}
\usepackage{adjustbox}

\def\sV{{\substack{\scalebox{0.6}{V}}}}
\def\sB{{\substack{\scalebox{0.6}{B}}}}

\def\selectr{{\substack{\scalebox{0.6}{elec}}}}  

\def\svib{{\substack{\scalebox{0.6}{vib}}}}   
\def\szero{{\substack{\scalebox{0.6}{0}}}}    

\def\VASP{{\small VASP}}
\def\SPBE{{\small PBE}}

\def\QHA{{\small QHA}}

\def\QHAPPP{{\small QHA3P}}

\def\GIBBS{{\small GIBBS}}

\def\EXP{{\mathrm{exp}}}

\definecolor{pranab_green}{rgb}{0.31,0.53,0.10}
\definecolor{pranab_red}{rgb}{0.85,0.23,0.11}

\makeatletter \renewcommand\frontmatter@abstractwidth{\dimexpr\textwidth\relax} \makeatother 

\makeatletter
 \def\@testdef #1#2#3{%
   \def\reserved@a{#3}\expandafter \ifx \csname #1@#2\endcsname
  \reserved@a  \else
 \typeout{^^Jlabel #2 changed:^^J%
 \meaning\reserved@a^^J%
 \expandafter\meaning\csname #1@#2\endcsname^^J}%
 \@tempswatrue \fi}

\begin{document}
\title{High-throughput screening of the thermoelastic properties of ultra-high temperature ceramics}

\author{Pinku Nath}
\affiliation{Lovely Professional University, India}
\author{Jose J. Plata}
\affiliation{Departamento de Química Física, Universidad de Sevilla, Seville, Spain}
\email[]{jplata@us.es}
\author{Julia Santana}
\affiliation{Departamento de Química Física, Universidad de Sevilla, Seville, Spain}
\author{Ernesto J. Blancas}
\affiliation{Departamento de Química Física, Universidad de Sevilla, Seville, Spain}
\author{Antonio M. Márquez}
\affiliation{Departamento de Química Física, Universidad de Sevilla, Seville, Spain}
\author{Javier Fdez. Sanz}
\affiliation{Departamento de Química Física, Universidad de Sevilla, Seville, Spain}
\email[]{jose@}

\date{\today}


\begin{abstract}
\noindent
Ultra-high temperature ceramics, UHTCs, are a group of materials with high technological interest because their use in extreme environments.
However, their characterization at high temperatures represents the main obstacle for their fast development.
Obstacles are found from a experimental point of view, where only few laboratories around the world have  the resources to test these materials under extreme conditions, and also from a theoretical point of view, where actual methods are extremely expensive.
Here, a new theoretical high-throughput framework for the prediction of the thermoelastic properties of materials is introduced. 
This approach can be systematically applied to any kind of crystalline material, drastically reducing the computational cost of previous methodologies.
Elastic constants for UHTCs have been calculated at a wide range of temperatures with excellent agreement with experimentally reported values.
Moreover, other mechanical properties such a bulk modulus, shear modulus or Poisson ration have been also explored.
Other frameworks with similar computational cost have been used only for predicting isotropic or averaged properties, however this new approach opens the door to the calculation of anisotropic properties at a very low computational cost.

\end{abstract}

\maketitle

\section{Introduction}

Ultra-high temperature ceramics, UHTCs, are usually defined as compounds whose melting point surpasses 3000\textdegree{}C~\cite{EWuchina_ESI_2007, UHTC_book, Fahrenholtz_sm_2017}. 
While UHTCs are not new materials for the scientific community and have been reported since late 1800s~\cite{Moissan_1896, Tucker_1902}, their technological interest started to grow in the late 1960s~\cite{Kalish_1969}. In the most recent decade, UHTCs have clearly emerged because of their potential use in extreme environments~\cite{UHTC_book}. 
Aerospace applications such as scramjet propulsion, hypersonic aerospace vehicles and advanced rocket motors are the main reason why research on UHTCs has grown in recent years~\cite{Levine_jecs_2002}. 
For instance, thermal control, mechanical resistance and corrosion are the main variables to consider when designing hypersonic vehicles, whose materials experience temperatures higher than 2000\textdegree{}C and are exposed to highly-reactive, dissociated gas species~\cite{VanWie_jms_2004}.
UTHCs combine high hardness, stiffness, and melting temperature with very low reactivity because of their strong covalent bonds between carbon, nitrogen or boron with transition metals, TM, such as Hf, Zr, Nb, Ti or Ta~\cite{Opeka_jecs_1999, Zhang_cms_2008}.

UHTCs-based materials have been rapidly developed during the last 25 years~\cite{Chamberlain_jacers_2004, Gasch_jms_2004,Tang_msea_2007,Monteverde_msea_2008}, but there are still many challenges to be tackled in order to spur the rational design, synthesis and deployment of these materials. 
The main issue that hampers the swift development of these materials is their experimental characterization and testing at extreme environments. 
Most well-known properties of these materials are obtained at ambient conditions and there are few laboratories around the world with the resources to test materials under extreme conditions~\cite{Han_cst_2008,Jackson_jacers_2011,Neuman_jacers_2013,Miller_jacers_2015,Paul_aac_2016}. 
Thus far, computational approaches have not presented solutions to these experimental barriers. 
Most of theoretical works related to UHTCs are focused on 0~K properties~\cite{Zhao_jssc_2008,Zhang_jac_2011,Zeng_prb_2013} and there are few reports in which temperature-dependent mechanical properties of UHTCs are explored~\cite{Cheng_jacers_2015,Xiang_jacers_2017}. 
This has been due to i) the lack of commercial algorithms to predict these temperature-dependent properties and ii) the high computational cost of these calculations.

There are methods, such as the quasi-harmonic approximation, \QHA~\cite{Carrier_prb_2007,Baroni_rmg_2010,curtarolo:art114}, that provide a relatively inexpensive computational approach to obtain temperature-dependent mechanical properties such as bulk or shear modulus.
However, {\QHA} is most frequently used with isotropic volume deformations of the crystal, so the properties obtained through this method can be considered as average mechanical features of the system.
That is the reason why elastic constants need to be computed in order to fairly capture the anisotropy of the material and obtain a more complete description of the temperature-dependent mechanical response of the system~\cite{Golesorkhtabar_ElaStic_CPC_2013,Zhao_prb_2007}.
Thus, elastic constants represent the starting point that  gives access to other mechanical properties.
Different approaches have been proposed to compute temperature-dependent elastic constants using {\QHA} as a formal framework~\cite{Davies_jpcs_1974,Karki_science_1999,Karki_prb_2000,Wu_prb_2011}.
However, their computational costs prevent them from being used systematically or routinely.
Other methods have been developed in order to reduce the cost of using the {\QHA} to compute temperature-dependent elastic constants~\cite{Kadas_prb_2007, Wang_jpcm_2010, Wang_jpcm_2010, Liu_prb_2019_2, Liu_cms_2019}.
For instance, the quasi-static approximation, QSA, reduces the number of calculations, assuming that the temperature dependence of the elastic constant is primarily due to thermal expansion~\cite{Kadas_prb_2007, Wang_jpcm_2010, Wang_jpcm_2010}.
Nevertheless, QSA tends to underestimate thermal effects and increase anisotropy, which is in detrimental to its use, especially at high temperatures~\cite{Destefanis_m_2019}. 

In this work, the elastic constants of UHTCs at finite temperatures are predicted to chart their mechanical properties, paying special attention to the high temperature range, in order to simulate their behavior at extreme conditions.
To do so, a new high-throughput framework has been designed that not only automatises the process, but also includes a new approach that reduces the computational cost compared with previous methodologies, without losing accuracy. 

\section{Methodology}

\subsection{Elastic constants}

Traditionally, elastic properties can be described within the Lagrangian theory of elasticity in which a solid is considered as a homogeneous and anisotropic elastic medium~\cite{ThermoCrys}. 
Within a linear regime and using the Voigt notation, the stress, \bm{$\sigma$}=($\sigma_1, \sigma_3, \sigma_3, \sigma_4, \sigma_5, \sigma_6$), 
and strain, \bm{$\epsilon$}=($\epsilon_1, \epsilon_2, \epsilon_2, \epsilon_2, \epsilon_2, \epsilon_6$), relation can be expressed as~\cite{Destefanis_m_2019,SHZhang_cpc_2017,GLUI_prb_2019}
\begin{equation}\label{stress_strain}
\sigma_i=\sum_{j=1}^6 c_{ij}\epsilon_j,
\end{equation}
where $c_{ij}$ are elastic stiffness constants of a crystal represented in a $6\times6$ matrix where $c_{ij}=c_{ji}$.
Considering this constraint, the total number of independent elastic components are 21 instead of 36.
Alternatively, it is possible to define the total energy of a crystal in terms of a power series of the strain~\cite{Golesorkhtabar_ElaStic_CPC_2013} as
\begin{equation}\label{energy_strain}
E({\bm \epsilon)} = E_0 + V_0 \sum_{i} \sigma_i^{(0)} \epsilon_i + \frac{V_0}{2!} \sum_{i,j} c_{ij} \epsilon_i \epsilon_j + ...
\end{equation}
where $E_0$ and $V_0$ are the DFT energy and volume of the reference structure.
If the optimized (ground state) structure is chosen as the reference, $\sigma_i^{(0)}= 0$  because equilibrium structure is stress free.

Two alternative expressions can be derived for the elastic constants according to Eqs. \ref{stress_strain} and \ref{energy_strain},
\begin{equation}\label{stress_strain2}
c_{ij} =  \frac{\partial \sigma_i}{\partial \epsilon_j} \biggm|_{\bm{\epsilon}=0} 
\end{equation}
and
\begin{equation}\label{energy_strain2}
c_{ij} =  \frac{1}{V_0} \frac{\partial ^2 E}{\partial \epsilon_i \partial \epsilon_j}\biggm|_{\bm{\epsilon}=0}. 
\end{equation}
Methods based on Eq.~\ref{stress_strain2} to calculate $c_{ij}$ are defined as "stress approach", while methods based on Eq.~\ref{energy_strain2} are classified as "energy approach".
Although both are based on the creation of strained structures, there are important differences between them.
The stress-strain approach is the most used method and a lower number of strained structures are needed~\cite{Nielsen_prl_1983,curtarolo:art100}. 
However, time-consuming calculations are required to obtain the same accuracy as with the results obtained with the energy-strain method using a less demanding setup~\cite{Zhao_prb_2007}. 
That is why the energy-strain method is preferred to reduce the sensitivity of the results with respect to the calculation setup. 
When the energy-strain method is used, a set of distortion or deformation modes are chosen depending on the crystal symmetry~\cite{Golesorkhtabar_ElaStic_CPC_2013}. 
For each distortion mode, different structures are created in which the amplitude or magnitude of the deformation is modified obtaining energy-strain curves. 
%
Elastic constants at $0$~K can be calculated using these curves. \medskip

\subsection{Free energy and the quasiharmonic approximation}

In \QHA, the total free energy, $F$, of a system is a function of volume, $V$, and temperature, $T$, and it is described as the sum of three terms (Eq.~\ref{Eq:F_energy}): i) the vibration less total energy at $0$ K, E$_\szero$, ii) the vibrational free energy, $F_\svib$ and iii) free energy due to thermal electronic excitations, $F_\selectr$~\cite{Liu_Cambridge_2016,Wang_ACTAMAT_2004, Duong_jap_2011}.
The first term of the equation can be computed with different {\it ab-initio} packages~\cite{vasp_prb1996,quantum_espresso_2009}.
The second term is obtained integrating over the phonon density of states, pDOS~\cite{pinku_prm_2019,curtarolo:art114,Togo_scrmat_2015}.
The last term of the equation is calculated integrating over the electronic density of states, DOS.
\begin{equation}\label{Eq:F_energy}
F\left(V,T\right)=E_\szero\left(V\right)+F_\svib\left(V,T\right)+F_\selectr\left(V,T\right).
\end{equation}

The calculation of $F_\svib$ at a given $V$ is performed using the harmonic approximation, where $F_\svib$ includes anharmonic effects in the form of volume-dependent phonon frequencies~\cite{pinku_prm_2019,curtarolo:art114,Duong_jap_2011,Wang_ACTAMAT_2004,Liu_Cambridge_2016},
\begin{widetext}
\begin{equation}\label{Eq:F_vib}
F_\svib\left(V,T\right)=\frac{1}{N_q}\sum_{{\bf q}, j}\left( \frac{\hbar\omega_{j}({\bf q})}{2}+k_{\sB}T{\rm ln}\left[1-\EXP\left(-\frac{\hbar\omega_{j}(\bf q)}{k_{\sB}T}  \right)\right] \right), 
\end{equation}
\end{widetext}
\noindent
where $\hbar$ and $k_\sB$ are the reduced Planck and Boltzmann constants, and $\omega_{j}(\bf q)$ is the volume-dependent phonon frequency for the wave vector, {\bf q}, and phonon branch index $j$.
$N_q$ is the total number of wave vectors.

For metals and narrow band-gap systems, the contribution of $F_\selectr \left(  V,T \right)$ to $F$ could be important and includes temperature-dependent contribution of the electrons to the internal energy, $U_\selectr \left(  V,T \right)$, and the electronic entropy, $S_\selectr \left(  V,T \right)$~\cite{pinku_prm_2019,Wang_ACTAMAT_2004,Arroyave_ACTAMAT_2005,Liu_Cambridge_2016}:
\begin{equation}\label{Eq:F_ele}
F_\selectr \left(V,T\right)= U_\selectr \left(V,T\right) - TS_\selectr \left(V,T\right).
\end{equation}
Both terms can be calculted as, 
\begin{equation}\label{Eq:U_ele}
U_\selectr \left(V,T\right)= \int_{0}^{\infty}  n_\selectr \left(\epsilon \right)f\left(\epsilon\right)\epsilon {\rm d}\epsilon - \int_{0}^{E_{\rm F}} n_\selectr \left(\epsilon\right)\epsilon {\rm d}\epsilon 
\end{equation}
and
\begin{equation}\label{Eq:S_ele}
\begin{split}
S_\selectr \left(V,T\right)=&-k_{\sB}\int_{0}^{\infty} n_\selectr \left(\epsilon \right) [f\left(\epsilon\right){\rm ln}\left(f\left(\epsilon\right)\right)+ \\ 
                            &+ \left(1-f\left(\epsilon\right)\right){\rm ln}\left(1-f\left(\epsilon\right)\right)]{\rm d}\epsilon
\end{split}
\end{equation}
where $n_\selectr(\epsilon)$ is the density of states at energy $\epsilon$, $f(\epsilon)$ is the Fermi-Dirac distribution, and $E_{\rm F}$ is the Fermi energy.

The temperature-dependent isothermal elastic constants, $c^T_{ij}(T)$, 
can be obtained by minimizing temperature dependent free energy, $F({\bm{\epsilon}},T)$ 
with respect to strain by using similar methodology as shown in Eq.~\ref{energy_strain2}.
The strain dependent $F({\bm{\epsilon}},T)$ at a given temperature, $T$ is defined as:
\begin{equation}\label{Eq:FE_energy}
F\left({\bm \epsilon},T\right)=E_\szero\left({\bm \epsilon}\right)+F_\svib\left({\bm \epsilon},T\right)+F_\selectr\left({\bm \epsilon},T\right), 
\end{equation}
where $F_\svib$ and $F_\selectr$ represent the vibrational and thermal electronic excitations of free energies. 
It is important to notice that $F_\svib\left({\bm \epsilon},T\right)$ and $F_\svib\left(V,T\right)$ are equivalent terms, because any strain applied to the material is directly connected to a specific volume.

\subsection{Accelerated \QHA\ (\QHAPPP)}
The $F_\svib$ in Eq. \ref{Eq:F_energy} is a function of $V$ and requires lattice dynamic calculations at several volumes.
This approach consumes large computational resources but it can be avoided with the use of the \QHAPPP\ method~\cite{pinku_prm_2019}.
In the \QHAPPP\ method, three phonon calculations are performed at three different volumes, one at the DFT
optimized volume, $V_\szero$, and the other two at $V_{\szero} \pm \Delta V$ volumes. 
$\Delta V$ represents small distorted volume from $V_\szero$.
The three obtained phonon frequencies are then used to extrapolate to any given volume at a given $\bf{q}$ using~\cite{Huang_cms_2016, pinku_prm_2019}:
\begin{equation}
\begin{split}
\omega\left({\bf q}, V\right)&=\omega \left({\bf q}, V_\szero\right)+\left(\frac{\partial \omega ({\bf q})}{\partial V}\right)_{\sV_{\szero}}(\Delta V)\\
      &+\frac{1}{2}\left(\frac{\partial^2\omega ({\bf q})}{\partial V^2}\right)_{\sV_{\szero}}\left(\Delta V\right)^2,
\end{split}
\label{Eq:4}
\end{equation}
The remaining steps to compute thermodynamic properties are the same as in the \QHA.

%

\subsection{Isothemal and adiabatic elastic constants}

As mentioned previously, isothermal elastic stiffness constants, $c^T_{ij}(T)$, can be calculated at finite temperatures by substituting internal Energy, $E$, by free energy, $F$, in Eq. \ref{energy_strain2},.
However, from an experimental point of view, elastic constants are generally obtained in adiabatic rather than isothermal conditions, using techniques such as ultrasonic measurements or Brilluoin scattering experiments~\cite{Featherston_pr_1963, Davenport_prb_1999}.  
Adiabatic elastic constants, $c^S_{ij}(T)$, are always equal to or larger than $c^T_{ij}(T)$.
In order to compare with experiments, $c^T_{ij}(T)$ are converted into $c^S_{ij}(T)$ following the relation reported by Davies\cite{Davies_jpcs_1974},
\begin{equation}
c^S_{ij} = c^T_{ij} + \frac{TV}{C_V} \lambda_i \lambda_j,
\label{Davies}
\end{equation}
where
\begin{equation}
\lambda_i = \sum_k \alpha_k c^T_{ik},
\label{Davies2}
\end{equation}
$\alpha_i$ is the linear thermal expansion coefficient in the direction $i$, $C_V$ is the specific heat and $\rho$ is the density.
For cubic systems~\cite{Destefanis_m_2019}: 
\begin{equation}
\lambda_1 = \lambda_2 = \alpha \left( c^T_{11} + c^T_{12} \right)
\label{Davies_cubic}
\end{equation}
and $\lambda_4 = 0 $ so $c^S_{44} = c^T_{44}$.
For hexagonal systems~\cite{Basaadat_ijmpc_2020},
\begin{equation}
\lambda_1 = \lambda_2 = \alpha_a \left( c^T_{11} + c^T_{12} \right) + \alpha_c c^T_{13}
\label{Davies_hex1}
\end{equation}
and 
\begin{equation}
\lambda_3 = 2\alpha_a c^T_{13} + \alpha_c c^T_{33},
\label{Davies_hex1}
\end{equation}
where $\alpha_a$ and $\alpha_c$ are he linear expansion coefficients in the directions "$a$" and "$c$". 

\subsection{Workflow and computational details}

\begin{figure*}[hbt!]
  \includegraphics[width=0.75\textwidth]{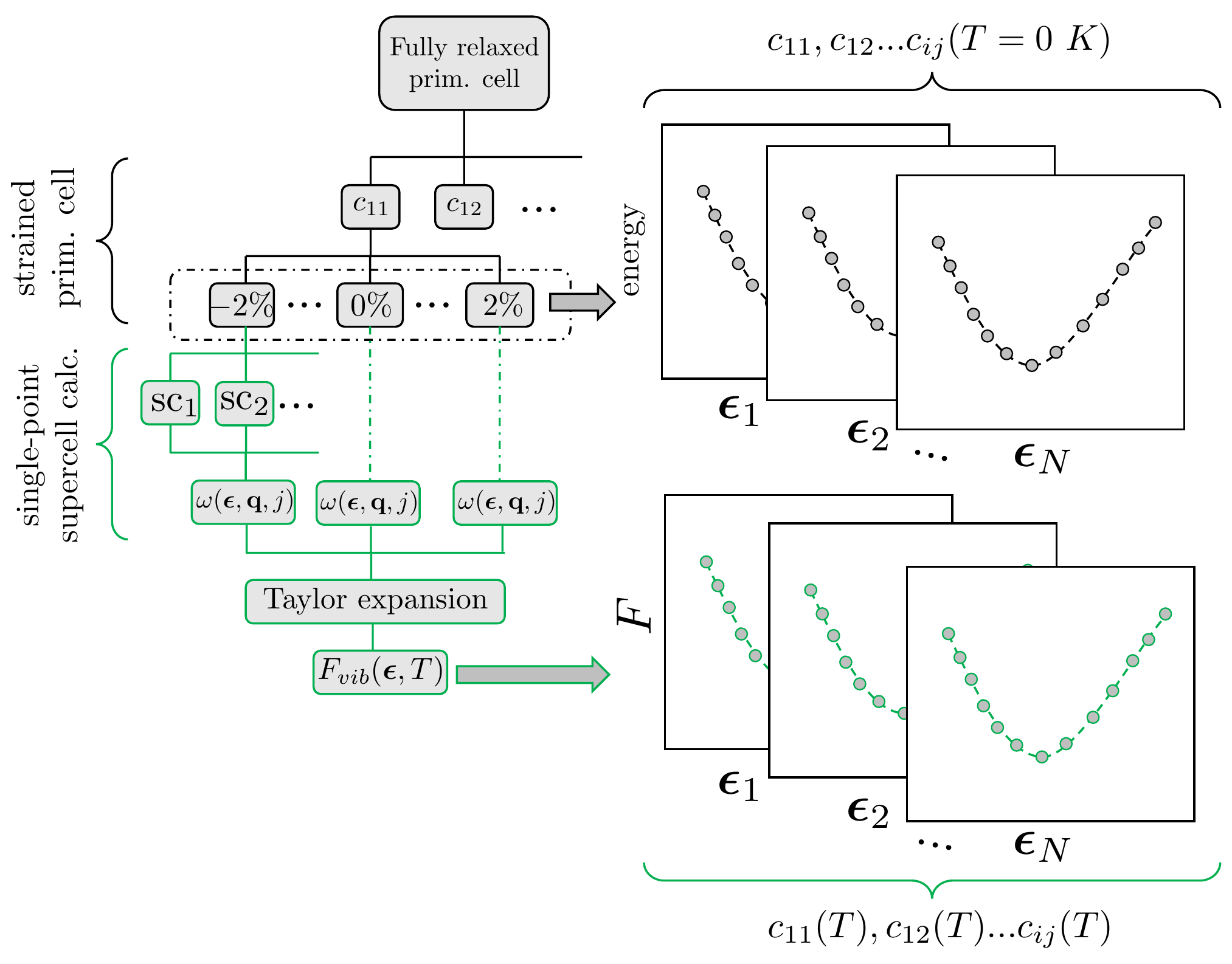}  \vspace{-2mm}
  \caption{\small 
  The workflows for calculating temperature dependent elastic constants.  
    }
  \label{fig:flowchart}
\end{figure*}

A high-throughput framework has been developed to automate the calculation of the temperature-dependent mechanical properties of UHTCs (Fig.~\ref{fig:flowchart}).
First, the primitive cell is fully optimized to characterize the minimum of the potential surface energy at 0~K.
The energy-strain method is used to calculate the elastic constants of each material so their space groups are calculated using spglib library~\cite{spglib_python} to identify the inequivalent $c_{ij}$.
Thirteen volume-distorted cells are generated for each inequivalent $c_{ij}$, 6 of them with a negative strain and 6 of them with a positive strain.
The set of distortion modes have been chosen following Zhang {\it et al.} approach~\cite{SHZhang_cpc_2017}. 
For instance, the distortions for cubic system are $\bm{ \epsilon}_1 = (0 , 0 , 0 , \delta , \delta, \delta ), \bm{ \epsilon}_{2} = ( \delta, \delta, 0 , 0 , 0 , 0)$ and $\bm{ \epsilon}_{3} = ( \delta, \delta, \delta, 0 , 0 , 0)$.
Where $\delta$ represents the magnitude of strain and goes from -2\% to +2\% of each lattice vector to obtain the 13 volume-distorted cells for each $\bm{ \epsilon}_i$.
In order to compute  $F({\bm{\epsilon}}, T)$, only three phonon calculations, including two distortions, $\delta$, and the fully-optimized geometry are required for a given $\bm{\epsilon}$. 
The frequencies obtained for these three structures are then used to estimate other frequencies for any arbitrary $\delta$ values for that particular $\bm{\epsilon}_i$ using a Taylor expansion.
Thus, for a cubic system a total of seven phonon calculations are required, including one phonon calculation at ${\bm \epsilon}=0$ state.
This approach is much less computationally demanding than \QHA\ where, approximately, $37$ phonon calculations are required ($13$ $\delta$ per ${\bm \epsilon}_i$).
Finally, $F(\bm{\epsilon}, T)$ are fitted with cubic polynomial to extract elastic constants \cite{Destefanis_m_2019}.

\noindent
{\bf Geometry optimization.}
All 0~K ground state structures were fully relaxed (atoms and lattice) using \VASP\ package~\cite{kresse_vasp,vasp_prb1996}.
Energies were obtained combining the projector-augmented wave ({\small PAW}) potentials~\cite{PAW} with the exchange-correlation functional proposed by Perdew-Burke-Ernzerhof (\SPBE)~\cite{PBE}. 
The number of valence electrons for each atom was selected following standards proposed by Calderon \textit{et al.}~\cite{Calderon_cms_2015}.
All calculations use a high-energy cut-off, 700~eV for borides and 550~eV for carbides and nitrides. 
Reciprocal space was explored using a dense {\bf k}-point mesh of 12,000 {\bf k}-points per reciprocal atom, approximately.
Wavefunction was converged self-consistently until the energy difference between two consecutive electronic steps was smaller than $10^{-9}$ eV.
Partial occupancies for each orbital were determined using Methfessel-Paxton method of order one.
Geometry optimizations were performed using three-atom primitive cell for borides and 8-atoms conventional cells for carbides and nitrides. 
Structures were considered fully relaxed when forces over all atoms were smaller than $10^{-8}$ eV/\AA.
An additional support grid for the evaluation of the augmentation charges was included to reduce the noise in the forces. 

\noindent {\bf Distorted cells.}
Elastic constants prediction requires three and five distortion modes for cubic and hexagonal systems, respectively.
These distortion modes are generated using the procedure shown in Ref. \cite{SHZhang_cpc_2017}.
Thirteen values of $\delta$ have been chosen within the range $\pm$ 2\%  for a ${\bm \epsilon}_i$.
These generated structures are relaxed without changing the cell volume.
%
%
%
%
%
%
After the relaxation, single point calculations were performed for each volume-distorted cell in order to calculate their density-of-states, DOS. 
 
\noindent
{\bf Phonon calculations.}
Different packages can be used to predict the vibrational spectra of solids, such as Alamode~\cite{Tadano_jpcm_2014}, APL-AAPL~\cite{curtarolo:art125}, PHON~\cite{Alfe_cpc_2009} or Phonopy~\cite{Togo_scrmat_2015}.
In this case, phonon calculations were performed combining Phonopy and \VASP\ to obtain second-order interatomic force constants, IFCs, via the finite displacement approach~\cite{curtarolo:art114}.
For each ${\bm \epsilon}_i$, two phonon calculations are performed at $\pm 2\%$ distortions including one at equilibrium structure.
Forces were extracted from $4\times4\times4$ supercells for borides (192 atoms) and $3\times3\times3$ supercells for nitrides and carbides (216 atoms).
The magnitude of the displacement to obtain the force constants was 0.01~\AA.
The same SCF convergence criteria followed in the optimizations was used for these calculations.
Frequencies and other related phonon properties such as $F_\svib$ were calculated using a $31\times31\times31$ \textbf{q}-point mesh that ensures their convergence. 

\section{Results}

\subsection{Phonon dispersion curves}

Phonon dispersion curves are an essential part to compute the vibrational contribution to the free energy and, simultaneously, give information about the stability of these materials. 
The absence of imaginary frequencies confirms the dynamic stability of UHTCs at 0~K (Fig.~S1).
Moreover, our results are in good agreement with previous theoretical predictions~\cite{Isaev_jap_2007}, and, most importantly, experimental data~\cite{Aizawa_prb_2001, Pintschovius_jpcss_1978, Kress_prb_1978, Weber_prb_1973, Christensen_prb_1979, Christensen_prb_1983, Smith_proc_1971, Smith_prl_1972}.
Only small deviations were found for the borides, which can be attributed to the measurement of the phonon dispersion curve at finite temperatures and specific surfaces~\cite{Aizawa_prb_2001}.  
     
\subsection{Elastic constants}

\begin{figure*}[htb!]
  \includegraphics[width=0.95\textwidth]{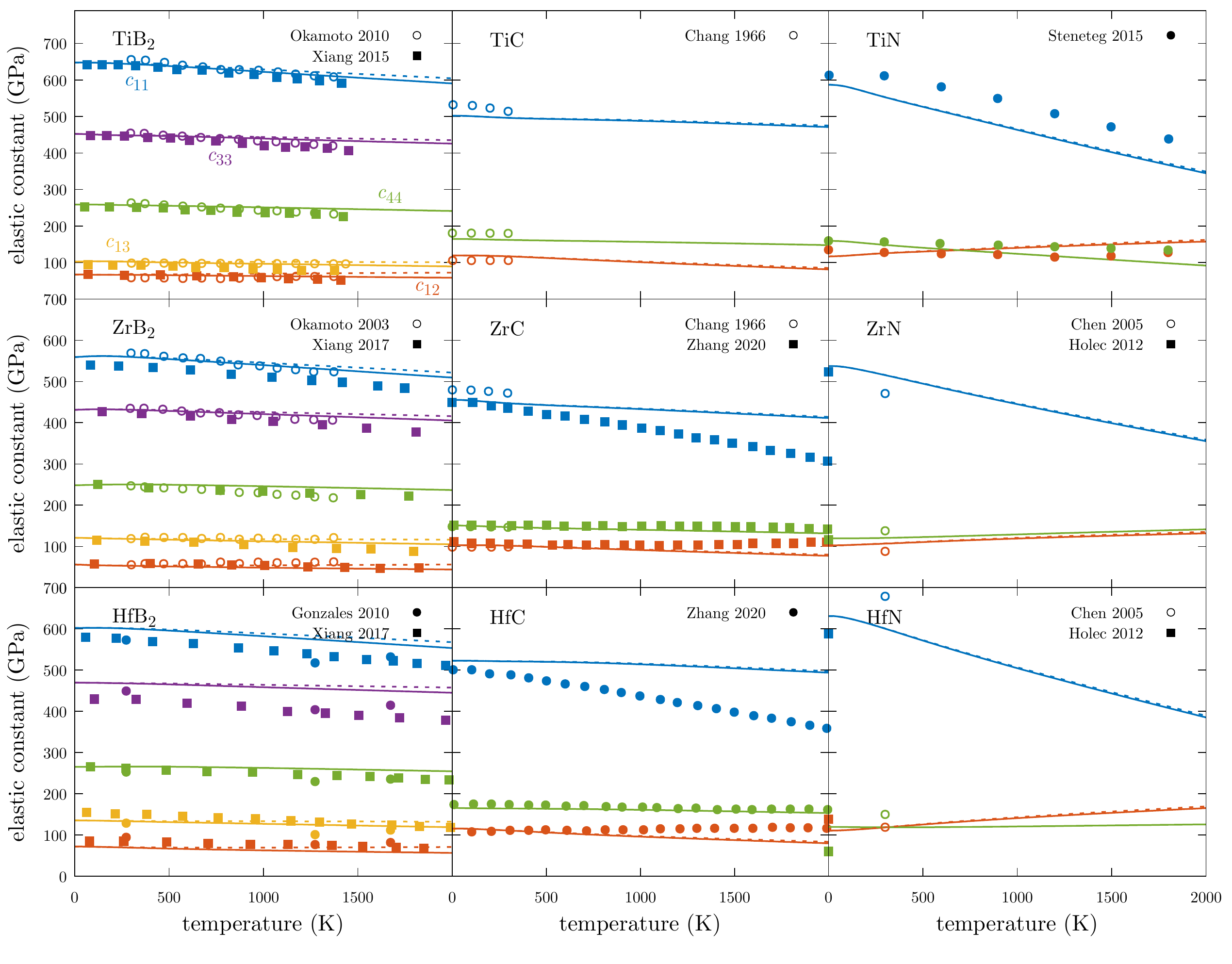}  \vspace{-2mm}
  \caption{\small 
    Isothermal (solid lines) and isentropic (dashed lines) elastic constants for UHTCs. Open points represent experimental measurements while filled points represent calculated values. Colors: $c_{11}$ = blue; $c_{12}$ = orange; $c_{13}$ = yellow; $c_{33}$ = purple; $c_{44}$ = green.}
  \label{fig:elastic}
\end{figure*}

In this section, calculated isothermal and isentropic elastic constants are compared with previous experimental values and simulations where available (Fig.~\ref{fig:elastic}).
To the best of our knowledge, TiB$_2$ and ZrB$_2$ are the only two UHTCs whose elastic constants have been experimentally well characterized in a wide range of temperatures~\cite{Okamoto_mrssp_2003,Okamoto_am_2010}.
The values obtained with the new high-throughput framework are in excellent agreement with the experiments and other calculated values obtained with more computationally demanding approaches~\cite{Xiang_jap_2015, Xiang_jacers_2017} (Fig. S2).
Experimental results, but in a shorter range of temperatures, were also found for TiC and ZrC~\cite{Chang_jap_1966}, with relative errors always below 5\%. 
Only room temperature values have been reported for ZrN and HfN~\cite{Chen_pnas_2005}, so it is difficult to analyze any trends.
That is why, we have also included calculated 0~K elastic constants for these two systems~\cite{Holec_prb_2012}. 
In both materials, calculated and experimental available results are aligned with our results and only c$_{11}$ at 300~K present a small deviation.
Only calculated values have been found for TiN~\cite{Steneteg_prb_2015} and HfC~\cite{zhang_arXiv_2020}.
Molecular dynamics performed by Steneteg {\it{et al.}} seem to predict similar trends and values than the HT framework for TiN.
For HfC, Zhang {\it{et al.}} obtained very similar values for $c_{12}$ and $c_{44}$ while $c_{11}$ seems to decrease faster than in our results.
However, they also obtained a very soft behavior of $c_{11}$ for ZrC~\cite{zhang_arXiv_2020}, while our methodology seems to follow the experimental measurements better. 
The fast reduction of $c_{11}$ with temperature is more noticeable for nitrides.
This trend is related to the changes in the volume when longitudinal strains are applied, which is strongly connected to temperature.
Other elastic constants such as $c_{12}$ and $c_{44}$ are related to deformation resistance to strain modes that are not connected to big changes in the volume, so they are less affected by temperature. 

\subsection{Mechanical stability}

Elastic constants describe the response of the crystal to external forces, so they play an important role determining their mechanical stability.
Mechanical stability has been extensively explored by different theoretical and computational works~\cite{Karki_jpcm_1997}.
Here, Born stability criteria~\cite{Born_Dynamic} will be adopted to elucidate the mechanical stability of these materials in a wide range of temperatures.
For a cubic crystal, the mechanical stability criteria~\cite{Wang_prl_1993} under isotropic pressure are,
\begin{equation}
\label{stability_cubic}
\begin{split}
&c_{44}>0, \\
&c_{11} - c_{12} > 0, \\
&c_{11} + 2 c_{12} > 0. \\
\end{split}
\end{equation}
For hexagonal systems, the stability criteria~\cite{Nye_Crystals, Mouhat_prb_2014} are:
\begin{equation}
\label{stability_cubic}
\begin{split}
&c_{44}>0, \\
&c_{11} - c_{12} > 0, \\
&\left( c_{11} + c_{12}\right) c_{33} - 2c^2_{13} > 0. \\
\end{split}
\end{equation}
All the materials explored in this work fulfil the Born stability criteria in the studied range of temperatures (Fig.~S3). 

\subsection{Isotropic mechanical properties}

Elastic constants are the essential ingredient to compute some key isotropic mechanical properties such as bulk modulus, $B$, shear modulus, $G$, Young's modulus $Y$, poisson ratio $\sigma$ and hardness $H_{\mathrm{V}}$.

\noindent
{\bf Bulk modulus and Shear.} Different definitions have been proposed to calculate the bulk modulus of an aggregate of crystals. 
Voigt´s definition is based on the averaging of the relation expressing the stress in a single crystal over all possible orientations~\cite{Voigt_book},
\begin{equation}
9B^X_{\mathrm{V}} =  \left( c^X_{11} + c^X_{22} + c^X_{33} \right) + 2 \left( c^X_{12} + c^X_{23} + c^X_{31} \right),
\label{BVoigt}
\end{equation}
where $X=(S,T)$ in order to differentiate between adiabatic and isothermal values, respectively.    
While Voigt´s definition assumes that the strain is uniform through the aggregate, Reuss´ approach consider that the stress is uniform~\cite{Reuss_ZAMM_1929},
\begin{equation}
\frac{1}{B^X_{\mathrm{R}}} = \left( s^X_{11} + s^X_{22} + s^X_{33} \right) + 2 \left( s^X_{12} + s^X_{23} + s^X_{31} \right),
\label{BReuss}
\end{equation}
where $\bm{s}^X$ is the compliance tensor, 
\begin{equation}
 \bm{s} = \bm{c}^{-1}. 
\label{compliance}
\end{equation}
It has been proven that Voigt moduli are always larger that Reus moduli with true values lying between them~\cite{Hill_pps_1952}. 
That is why, the Voigt-Reuss-Hill bulk modulus, $B^X_{\mathrm{VRH}}$, is defined as,
\begin{equation}
B^X_{\mathrm{VRH}} = \frac{1}{2} \left( B^X_{\mathrm{V}} + B^X_{\mathrm{R}}\right). 
\label{B_VRH}
\end{equation}
Similarly, Shear modulus can be defined as,
\begin{equation}
\begin{split}
G^X_{\mathrm{V}} =& \frac{1}{15} \left( c^X_{11} + c^X_{22} + c^X_{33} \right) -  \left( c^X_{12} + c^X_{23} + c^X_{31} \right) +\\
                  &+ 3\left( c^X_{44} + c^X_{55} + c^X_{66} \right)
\end{split}
\label{GVoigt}
\end{equation} 
or
\begin{equation}
\begin{split}
15/G^X_{\mathrm{R}} =& 4\left( s^X_{11} + s^X_{22} + s^X_{33} \right) - 4 \left( s^X_{12} + s^X_{23} + s^X_{31} \right) + \\
                     & + 3 \left( s^X_{44} + s^X_{55} + s^X_{66} \right).
\end{split}
\label{GReuss}
\end{equation}
Again, $G^X_{\mathrm{R}}<G^X_{\mathrm{V}}$ and real values should lie in between so,
\begin{equation}
G^X_{\mathrm{VRH}} = \frac{1}{2} \left( G^X_{\mathrm{V}} + G^X_{\mathrm{R}}\right). 
\label{G_VRH}
\end{equation}
Taking into account the small difference between isentropic and isothermal elastic constants and, in order to simplify the analysis of the results, $B^T$ and $G^T$ will be used to compute the other properties presented in this work.

\begin{figure*}[hbtp!]
  \includegraphics[width=0.95\textwidth]{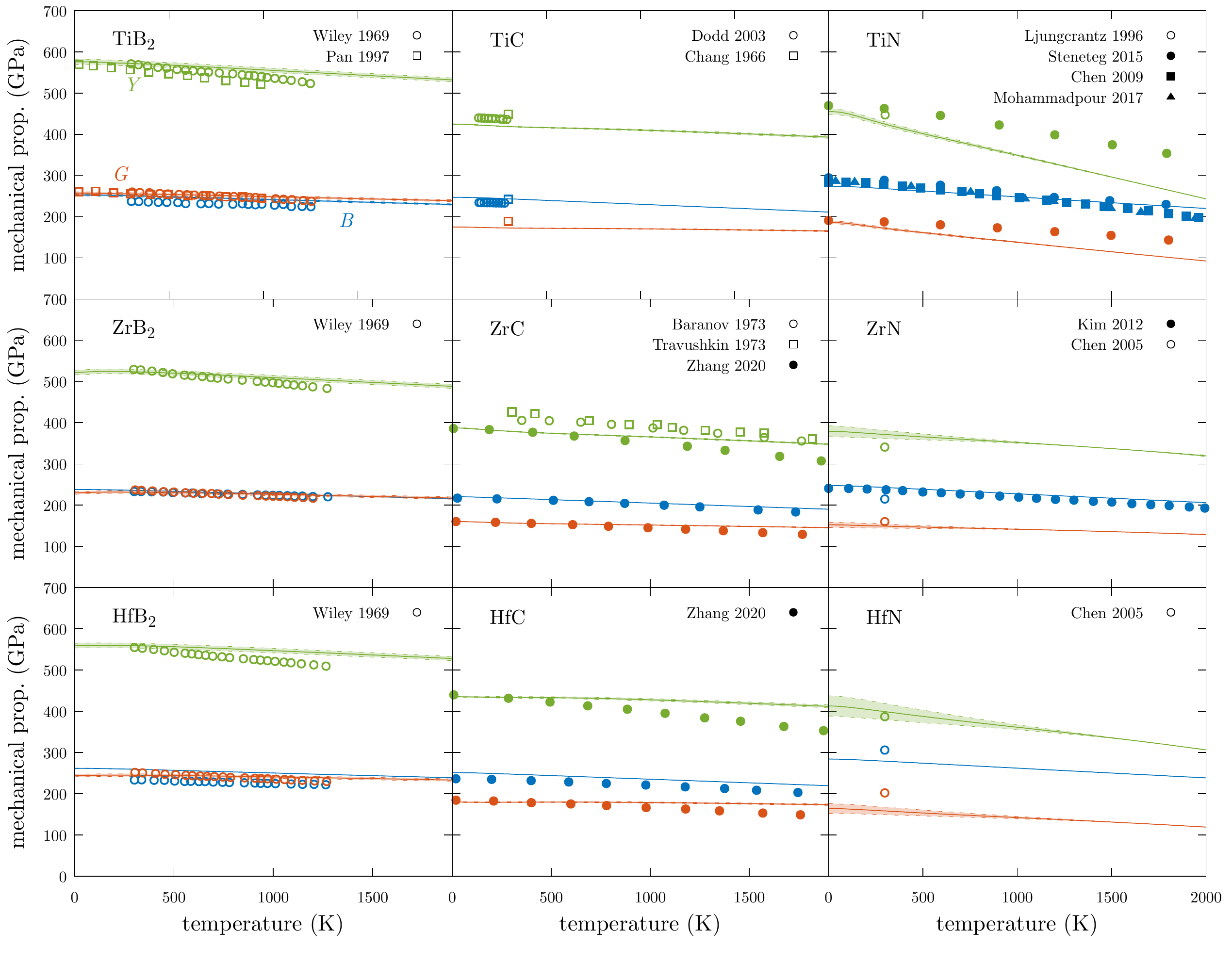}  \vspace{-2mm}
  \caption{\small 
  Bulk modulus, $B$ (blue), Shear modulus, $G$ (orange), and Young's modulus, $Y$ (green), for UHTCs. Voigt and Reuss values are depicted with dashed lines and Voigt-Reuss-Hill values are depicted with a solid line. The area ranged between Voigt and Reuss values has also been filled with the same color than the property. Open points represent experimental measurements while filled points represent calculated values.
 }
  \label{fig:BGE}
\end{figure*}

The comparison of simulations with experimental values for $B$ and $G$ is not a simple task (see Fig.~\ref{fig:BGE}). 
Most of the time, experimental measurements are obtained from polycrystalline samples in which porosity plays an important role, modifying their mechanical properties.
There are different models that correlate the mechanical properties of the fully dense material and the porosity with the mechanical properties of actual sample~\cite{Spriggs_jacers_1961, Nanjangud_jacers_1995, Rice_kem_1996}.  
Here, the Gibson and Ashby equation~\cite{Gibson_prsla_1982} was adopted to compare the experimental $B$, $G$ and Young's modulus, $Y$, with the theoretical results, if the porosity of the sample was reported.
For instance, predicted values are in excellent agreement with experimental measurements~\cite{Wiley_jlcm_1969,Pan_jacers_1997} for the $B$ and $G$ of borides, once porosity is considered. 
A similar trend is also found for TiC where experimental values are also available~\cite{Chang_jap_1966,Dodd_jms_2003}.
Larger deviations with respect to experimental values are obtained for HfN where $G$ is underestimated around 15\% at 298~K. 
Comparing with other previous theoretical results also helps to demonstrate how this new approach can be not only accurate but also how it can substantially reduce the computational time. 
For instance, the values obtained for $B$ and $G$ match well with the results reported by methods based on the \QHA\ in which phonon calculations are performed for all distorted structures for ZrC and HfC~\cite{zhang_arXiv_2020}, TiN~\cite{Chen_amse_2009,Mohammadpour_ms_2018} and ZrN~\cite{Kim_jac_2012}.

\noindent
{\bf Young's modulus and Poisson ratio.}
Young's modulus, $Y$, is a directional property, however, it can be initially assumed isotropic in order to extract a single approximate value for each compound,
\begin{equation}
 \label{young}
  Y=2G(1+\sigma),
 \end{equation}
where $\sigma$ is the isotropic Poisson ratio,
\begin{equation}
 \label{young}
  \sigma = \frac{\left( 3B - 2G\right)}{\left( 6B + 2G\right)}.
 \end{equation}
Similarly to $B$ and $G$, our calculations are in good agreement with experimental values, when available for $Y$ (Fig.~\ref{fig:BGE}).
Calculated values for borides~\cite{Wiley_jlcm_1969,Pan_jacers_1997} present a maximum relative error around 7\% at high temperatures, which is very small considering: i) the values are obtained using a very simple model to take into account the porosity of the sample~\cite{Spoor_apl_1997,Munro_2000} and ii) high-order force constants, which are not calculated here, can play an important role at high temperatures.
Similar trends are observed for TiC~\cite{Chang_jap_1966,Dodd_jms_2003} and ZrC~\cite{Baranov_sm_1973,Travushkin_sm_1973}, where predicted $Y$ values are slightly underestimated, but follow the same trend as experimental measurements.
If experimental reports were not available, previous theoretical works were used to evaluate the results obtained with this new high-throughput approach.
No significant discrepancies were found when $Y$ values were compared with the results obtained with methods based on the \QHA\ (see ZrC and HfC~\cite{zhang_arXiv_2020}).
Larger differences were found for TiN where Steneteg {\it{et al.}} used molecular dynamics to study the mechanical properties of TiN~\cite{Steneteg_prb_2015}.
However, $Y$ experimental values for TiN single crystals at room temperature are between 445-449~GPa~\cite{Ljungcrantz_jap_1996}, which are close to the values calculated in this work.    

To the best of our knowledge, there are not many experimental studies of the temperature dependence of the Poisson ratio.
For instance, Wiley {\it et al.} explored the Poisson ratio of the borides of the group IV up to 1300~K~\cite{Wiley_jlcm_1969}.
Our results are not only in agreement at room temperature, but also reproduce the very small variation that this property presents in a large temperature range (Table~\ref{tab:poisson}).
When compared with borides, the temperature dependence is slightly higher for carbides but is even larger in TiN and HfN.
This trend is also observed in the wider range of experimental values previously reported (Table~\ref{tab:poisson}).  

\begin{table*}
  \caption{\small 
  Comparison of the calculated Poisson ratio for UHTCs in this work in a 0-2000~K temperature range and experimental reported values (exp.). 
 }
\begin{tabular}{@{}ccccccc@{}}
\toprule
\multicolumn{1}{c}{} & \multicolumn{2}{c}{B} & \multicolumn{2}{c}{C} & \multicolumn{2}{c}{N}\\
\multicolumn{1}{c}{} & this work & exp. & this work & exp. & this work & exp. \\
\cmidrule{2-7}
Ti   & 0.11-0.12 & 0.10-0.11~\cite{Wiley_jlcm_1969} & 0.18-0.21 & 0.17-0.19~\cite{Kral_Lengauer_1998}; & 0.22-0.36 & 0.30~\cite{Kral_Lengauer_1998}; \\
     &           &                                  &           & 0.19~\cite{Yang_jac_2000}            &           & 0.22~\cite{Yang_jac_2000} \\
Zr   & 0.12-0.13 & 0.11-0.12~\cite{Wiley_jlcm_1969} & 0.18-0.21 & 0.19-0.26~\cite{Kral_Lengauer_1998}; & 0.23-0.25 & 0.19-0.25~\cite{Kral_Lengauer_1998};\\
     &           &                                  &           & 0.20~\cite{Yang_jac_2000}            &           & 0.26~\cite{Yang_jac_2000}\\
Hf   & 0.13-0.14 & 0.12-0.13~\cite{Wiley_jlcm_1969} & 0.18-0.21 & 0.16-0.18~\cite{Kral_Lengauer_1998}; & 0.24-0.32 & 0.26-0.35~\cite{Kral_Lengauer_1998};\\ 
     &           &                                  &           & 0.16~\cite{Yang_jac_2000}            &           & 0.17~\cite{Yang_jac_2000}\\
\bottomrule
\end{tabular}
  \label{tab:poisson}
\end{table*}

\begin{figure}[h!]
  \includegraphics[width=0.95\columnwidth]{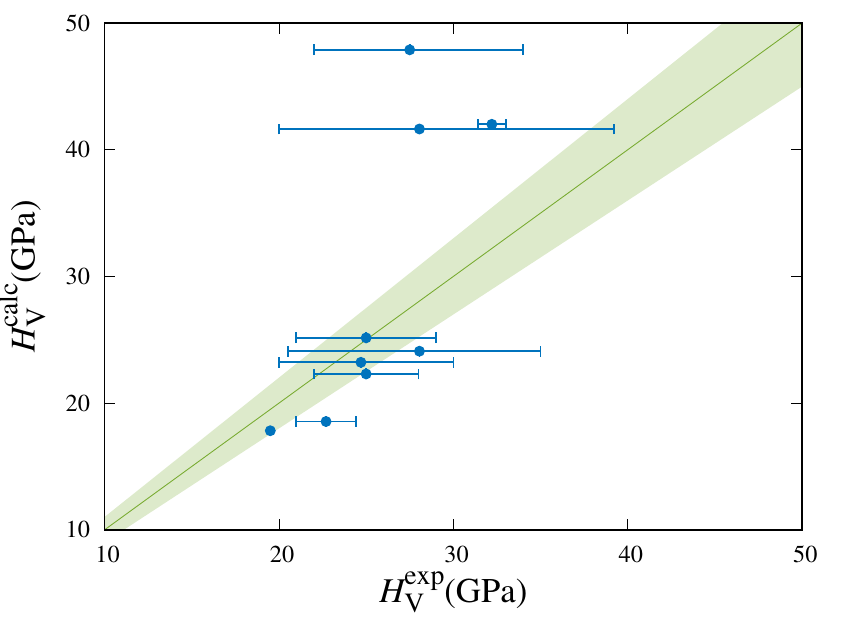}  \vspace{-2mm}
  \caption{\small 
  Comparison between calculated, $H_{\mathrm{V}}^{\mathrm{calc}}$, and experimental, $H_{\mathrm{V}}^{\mathrm{exp}}$, hardness.  
  Green area represent the $\mp$10\% relative error with respect to experimental values.
 }
  \label{fig:hv300}
\end{figure}

\noindent {\bf Hardness.}
Hardness is probably one of the most difficult mechanical properties to predict and compare with experimental data. 
It not only presents a dependency with the porosity~\cite{Yu_ijrmh_1999} or grain size~\cite{Rice_jacers_1994}, but also with other variables related to the measurement, such as indentation type (nano or micro), load and time~\cite{Gong_msea_2001,Guo_jmpt_2018}.
During the last two decades, different models have been proposed to predict the hardness of materials~\cite{Gao_prl_2003, Simunek_prl_2006, Chen_intermetallics_2011}.
Most of them assume isotropic conditions which could  overestimate the hardness of some materials, depending on the crystal plane exposed on the surface.~\cite{Chen_prb_2011}.
Moreover, porosity and indentation load tend to reduce the values obtained for hardness~\cite{Munro_2000}.
Here, hardness is calculated using the approach proposed by Tian \textit{et al.}~\cite{Tian_refmat_2012},
\begin{equation}
 \label{hardness}
  H_{\mathrm{V}}=0.92k^{1.137}G^{0.708},
 \end{equation}
where $k$ is the Pugh's modulus, which is defined as the ratio between the shear, $G$, and the bulk modulus, $B$.
In Eq. \ref{hardness}, the constants are adjusted to obtain  $H_{\mathrm{V}}$ in GPa units.
At room temperature, borides seem to be the most overstimated values obtaining 47~GPa, 41~GPa and 42~GPa for Ti$B_2$, Zr$B_2$ and HfB$_2$ when experimental values range between 34-22~GPa~\cite{Shackelford_CRC, Munro_2000, Otani_jcsj_2000, Raju_jecs_2009}, 39-20~GPa~\cite{Fahrenholtz_jacers_2007, Xuan_jpd_2002, Csanadi_jecs_2016} and 33-31.4~GPa~\cite{Bsenko_jlcm_1974, Liang_ci_2019}, respectively (Fig.~\ref{fig:hv300}). 
TiC stands as a good example of the different values that can be obtained for hardness, depending on the plane and orientation of the crystal~\cite{Kumashiro_jms_1977, Maerky_msea_1996, Yang_jac_2000}.
For instance, TiC (100) plane on the $\left[110\right]$ direction presents a micro-Vickers hardness of 34.9~GPa while the values for the $\left(110\right)$ plane at the $\left[100\right]$ direction is 23.24~GPa~\cite{Kumashiro_jms_1977}.  
Using the approach proposed by Tian, the calculated value (24.1~GPa) is in that range experimentally reported at room temperature.
Similarly, calculated hardness for ZrC (23.2~GPa) and HfC (25.2~GPa) are in good agreement with the experimental measurements~\cite{Kohlstedt_jms_1973, Kumashiro_jmsl_1982, Balko_jecs_2017, Cheng_jecs_2017, ASM_book}.
To the best of our knowledge, there are not too many experimental works which study the hardness of nitrides, however we have found that calculated values are slightly lower than the homolog carbides and are close to the reported values for TiN~\cite{Andrievski_nanom_1997}, ZrN~\cite{Mei_jvsta_2013} and HfN~\cite{Chen_pnas_2005}.  

If predicting hardness is a difficult task because of the wide range of experimental variables, capturing the temperature dependence of this property is even more of a challenge.
Hardness changes with temperature.
Especially at high temperatures, hardness is controlled by creep due to dislocation diffusion phenomena.
The activation energy for creep can be calculated from
\begin{equation}
 \label{HvT}
   H_{\mathrm{V}}^{-m} = A \exp(-Q/RT) t
 \end{equation}
where $Q$, $R$, $T$, and $t$ are the activation  energy for creep,  the gas  constant, the  temperature  and  loading  time,  respectively,  and m and A are  constants~\cite{Atkins_prs_1966,Atkins_jim_1966}. 
When Arrhenius plots are used to study hardness in a temperature range, different regions are identified for many UHTCs~\cite{Kumashiro_jms_1977, Chen_jac_2003, Kohlstedt_jms_1973, Xuan_jpd_2002, Otani_jcsj_2000}.
These regions are linked to a brittle-ductile transition and the different mechanisms that govern the deformation during the indentation at different temperatures~\cite{Kohlstedt_jms_1973}.
Activation energies for creep for borides and carbides have been reported to be around 80~kcal/mol and most of experimental studies obtain values for $m$ between 4 and 5~\cite{Kohlstedt_jms_1973,Kumashiro_jms_1977}.
These values can be combined with our predictions at room temperature to calculate the pre-exponential constant and obtain good trends for hardness below 1000~K. 
 
\subsection{Anisotropy}
Dislocations dynamics or phase transformations are some examples in which anisotropy plays a relevant role~\cite{Ledbetter_jap_2006}.
Quantification of crystal anisotropy has been a subject of debate since Zener introduced the first anisotropy index~\cite{Zener_book}.
Here, the universal elastic anisotropy index~\cite{Ranganathan_prl_2008},
\begin{equation}
 \label{AU}
  A{^\mathrm{U}} =  \left( \frac{B_{\mathrm{V}}}{B_{\mathrm{R}}} \right) + 5\left( \frac{G_{\mathrm{V}}}{G_{\mathrm{R}}} \right) -6,
\end{equation}
and the log-Euclidean anisotropy index~\cite{Kube_aipa_2016},
\begin{equation}
 \label{AL}
  A{^\mathrm{L}} = \sqrt{ \left[ \ln \left( \frac{B_{\mathrm{V}}}{B_{\mathrm{R}}} \right) \right]^2 + 5 \left[ \ln \left( \frac{G_{\mathrm{V}}}{G_{\mathrm{R}}} \right)\right]^2},
\end{equation}
are used to calculate the temperature-dependent anisotropy of UHTs (Table~\ref{tab:ani_ind}).

\begin{table}[h!]
  \caption{\small 
  Anisotroy indexes $A{^\mathrm{U}}$ and $A{^\mathrm{L}}$ for UHTCs in the  0-2000~K temperature range. 
 }
\begin{tabular}{@{}ccccccc@{}}
\toprule
\multicolumn{1}{c}{} & \multicolumn{2}{c}{B} & \multicolumn{2}{c}{C} & \multicolumn{2}{c}{N}\\
\multicolumn{1}{c}{} & $A{^\mathrm{U}}$ & $A{^\mathrm{L}}$ & $A{^\mathrm{U}}$ & $A{^\mathrm{L}}$ & $A{^\mathrm{U}}$ & $A{^\mathrm{L}}$ \\
\cmidrule{2-7}
Ti   & 0.13-0.11 & 0.05-0.04 & 0.03-0.09 & 0.01-0.04 & 0.19-0.00 & 0.08-0.00 \\
Zr   & 0.15-0.14 & 0.06      & 0.03-0.07 & 0.01-0.03 & 0.44-0.07 & 0.19-0.03 \\
Hf   & 0.14-0.13 & 0.06      & 0.05-0.11 & 0.02-0.05 & 0.76-0.02 & 0.32-0.01 \\
\bottomrule
\end{tabular}
  \label{tab:ani_ind}
\end{table}

Borides, carbides and nitrides, in general, present relatively low anisotropy indexes if they are compared with other materials~\cite{Kube_aipa_2016}.
While the reported $A{^\mathrm{U}}$ and $A{^\mathrm{L}}$ for some cubic systems can be as low as 10$^{-3}$, some triclinic and monoclinic materials present $A{^\mathrm{U}}$ values higher than 10$^{2}$~\cite{Kube_aipa_2016}.
Both indexes point to the nitrides as the most anisotropic materials, followed by the borides and finally the carbides (Table~\ref{tab:ani_ind}).
These indexes are usually obtained from 0~K calculated mechanical properties, however, their behaviour at high temperatures can be calculated using our approach.  
For instance, anisotropic indexes slightly increase with temperature for carbides, while they are reduced for borides and nitrides.
Borides experiment change very little in their anisotropic indexes in the 0-2000~K range, however, both $A{^\mathrm{U}}$ and $A{^\mathrm{L}}$ are drastically reduced for nitrides. 

\begin{figure*}[hbtp!]
  \includegraphics[width=0.95\textwidth]{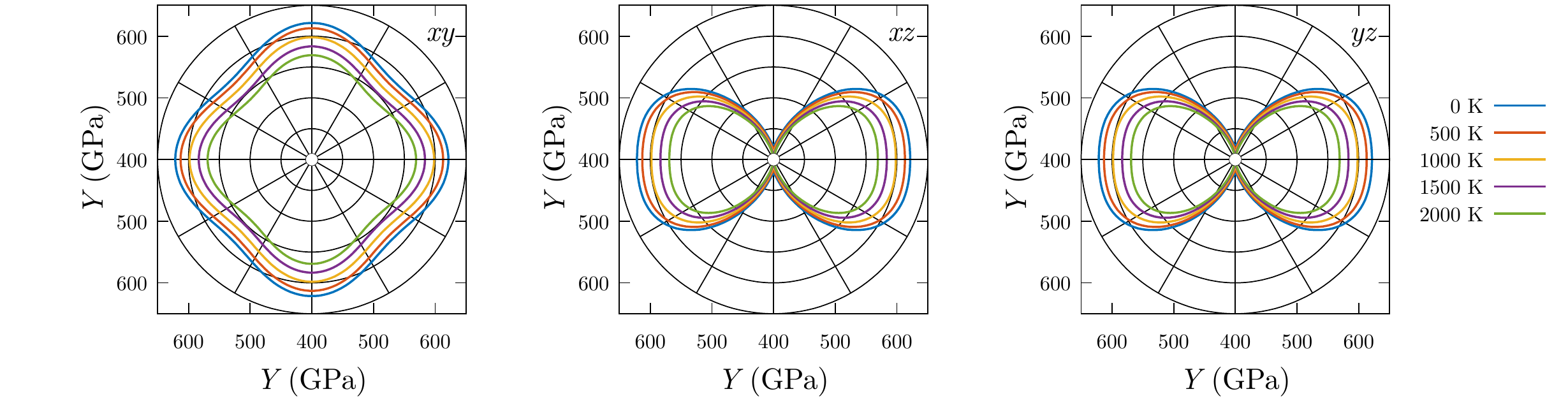}  \vspace{-2mm}
  \caption{\small
  Anisotropic behavior of $Y$ for TiB$_2$ at different temperatures. Left, mid and right panels correspond to $xy$, $xz$, and $yz$ planes, respectively.
 }
  \label{fig:aniTiB2}
\end{figure*}

Anisotropy indexes give relevant information about the general behavior of the material, however, some properties such as $Y$, $G$ and $\sigma$ can be directly calculated as a function of the crystallographic direction.
This information is extremely valuable in order to predict the behaviour of single crystal materials in specific direction or the limits of these properties in polycrystalline samples.
There are already different packages that calculate and represent the anisotropic nature of some mechanical properties~\cite{Marmier_cpc_2010,Ortiz_jcp_2013}. 
In this work, ELATE package~\cite{Gaillac_jpcm_2016} has be combined with our framework to explore the temperature-dependent anisotropic nature of some mechanical properties.
As an example, the directional and temperature dependence of $Y$ is plotted for TiB$_2$ in Fig.~\ref{fig:aniTiB2}, where $Y$ is a 33\% lower in $c$ with respect to $a$ and $b$

\begin{figure*}[hbtp!]
  \includegraphics[width=0.95\textwidth]{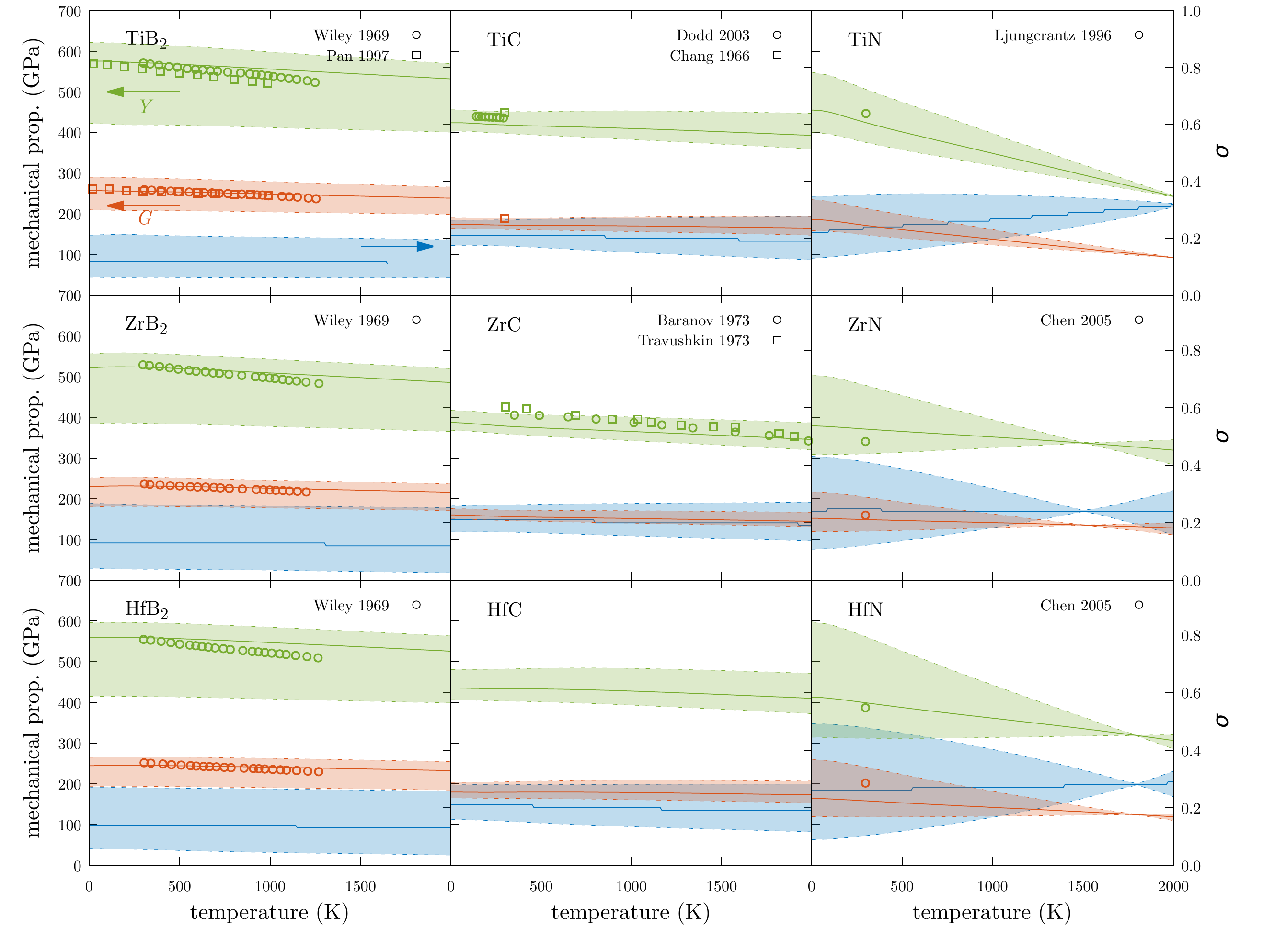}  \vspace{-2mm}
  \caption{\small
  Anisotropic Poisson ratio, $\sigma$ (blue), Shear modulus, $G$ (orange), and Young's modulus, $Y$ (green), for UHTCs. Solid lines represent isotropic values calculated in previous section. Dashed lines represent the upper and lower limit for each property. The area ranged between upper and lower limits has also been filled with the same color than the property. Open points represent experimental measurements.
 }
  \label{fig:ani}
\end{figure*}

The same approach has been followed for the rest of UHTCs materials studied in this work, not only for $Y$, but also for $G$ and $\sigma$ (Fig.~\ref{fig:ani}). 
The colored area for each property at each temperature is delimited by the maximum and minimum values (dashed lines) predicted for $Y$ (green), $G$ (orange) and $\sigma$ (blue).
Results in Fig.~\ref{fig:ani} are in good agreement with the trends extracted from anisotropy indexes.
Carbides are the group of materials with a lower variability in their properties, which correspond to lower anisotropy indexes than borides and nitrides.
Moreover, the difference between the maximum and minimum values in carbides slightly increases with temperature, following the same trend as the anisotropic indexes.
Similarly to values in Table~\ref{tab:ani_ind}, the amplitude between maximum and minimum for each property in borides remains almost constant with temperature, while a fast reduction of the amplitude can be observed for nitrides. 
Some singular points can be observed for nitrides, where maximum and minimum values are the same at a given temperature.
These points represent an inversion in the direction where maximum and minimum values of a specific property can be observed. 
For instance, $Y$ maximum values is observed in the $\left[100\right]$ direction up to 1500~K approximately.
For temperatures higher than 1500~K, this trend is different and the minimum value for $Y$ is obtained in $\left[100\right]$ direction.
Same phenomena is observed in TiN and HfN  around 2000~K and 1800~K, respectively.
Fig.~\ref{fig:ani} is also a good approach to visualize the potential scattering in experimental measurements depending on the crystallinity, direction and temperature in which the property has been measured.
Experimental data already plotted in Fig.~\ref{fig:BGE} is always in between the maximum and minimum limits stablished for each property at each temperature.
Only one experimental point is slightly out of the delimited area for ZrC.
This very small deviation could be due to the well known underestimation of bond strength by GGA functionals.
    
\subsection{Thermodynamic and thermal properties}

\begin{figure*}[hbtp!]
  \includegraphics[width=0.95\textwidth]{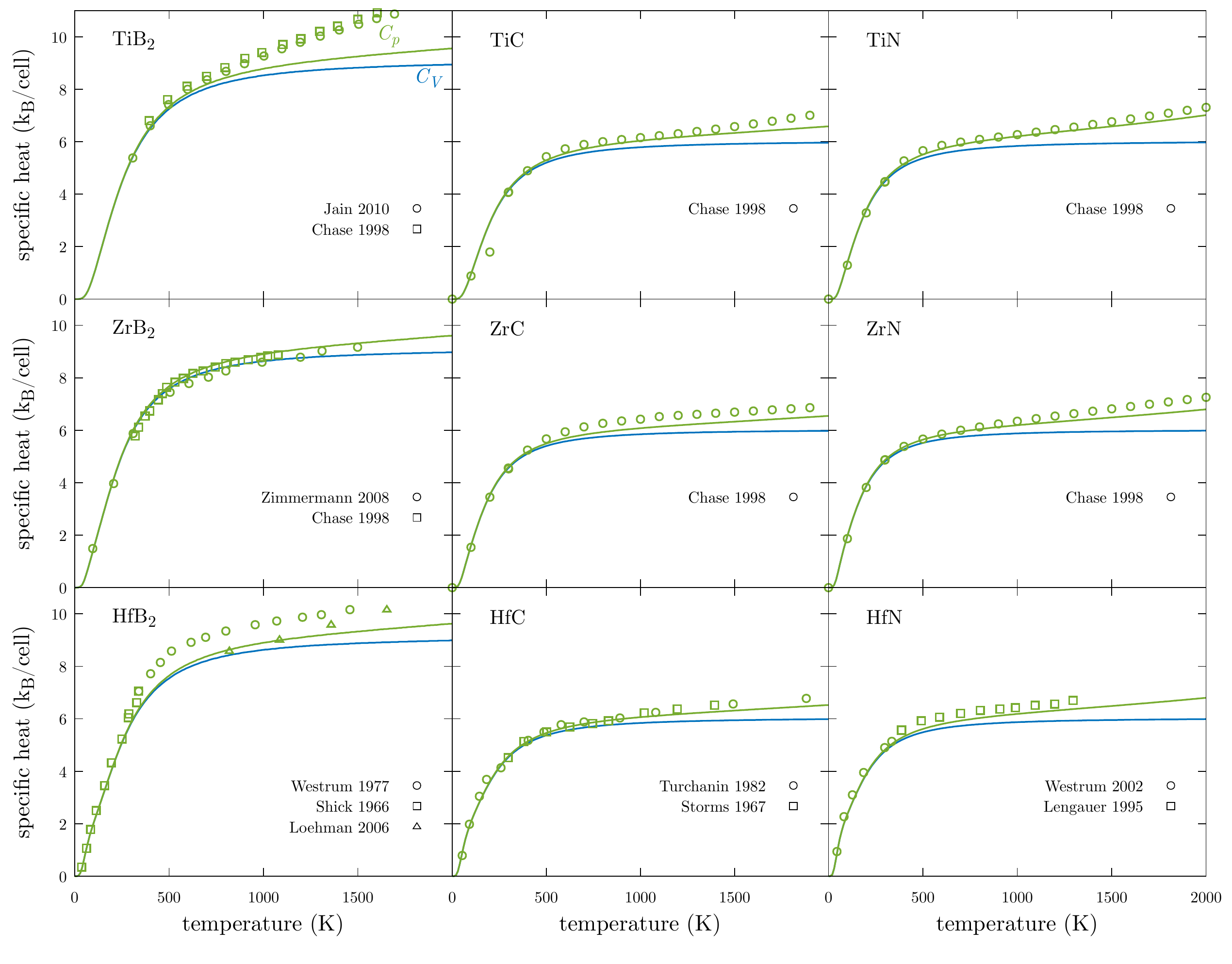}  \vspace{-2mm}
  \caption{\small
   Heat capacity at constant volume, $C_V$ (blue), and at constant pressure, $C_p$ (green) for UHTCs. Solid lines represent calculated values while open points represent experimental measurements.
 }
  \label{fig:heat}
\end{figure*}

\noindent \textbf{Heat capacity}. Thermodynamic properties such as specific heat at constant volume, $C_V$, are calculated including phonon,  $C^{\mathrm{ph}}_V$, and electronic, $C^{\mathrm{el}}_V$, contributions:
\begin{equation}
 \label{cv}
  C_V = C^{\mathrm{ph}}_V + C^{\mathrm{el}}_V.
\end{equation}
Phonon contribution is calculated as,
\begin{equation}
 \label{cv_p}
  C^{\mathrm{ph}}_V = \frac{k_{\mathrm{B}}}{N_q}\sum_{{\bf q}, j} c_{{\bf q}, j}, 
\end{equation}
where
\begin{equation}
 \label{cv_p2}
  c_{{\bf q}, j}   = \left(\frac{\hbar \omega_{{\bf q}, j}(V_0)}{k_{\mathrm{B}}T}\right)^2\frac{\exp\left(\frac{\hbar \omega_{{\bf q}, j}(V_0)}{k_{\mathrm{B}}T}\right)}{\left(\exp\left(\frac{\hbar \omega_{{\bf q}, j}(V_0)}{k_{\mathrm{B}}T}\right)-1\right)^2}.
\end{equation}
Following the free electron gas approximation~\cite{ashcroft}, the electronic contribution to heat capacity is a linear function with respect to the temperature,
\begin{equation}
 \label{cv_e}
  C^{\mathrm{el}}_V = \frac{1}{3}\pi^2N(E_{\mathrm{F}})k_{\mathrm{B}}^2T, 
\end{equation}
where $N(E_{\mathrm{F}})$ is the density of states (DOS) at the Fermi level.
Moreover, the specific heat at constant pressure,$C_p$, which is more experimentally accesible, was calculated as~\cite{Grimvall_1999},
\begin{equation}
 \label{cp}
C_p  = C_V+ V_{\rm{eq}}TB\alpha_V^2,
\end{equation}
where $V_{\rm{eq}}$ is the equilibrium volumen for a given temperature and $\alpha_V$ is the volumetric lattice thermal expansion.
In most cases, specific heats match experimental values well~\cite{Jain_jac_2010,Chase_AIP_1998,Zimmermann_jacers_2008,Westrum_jct_1977,TCRC_book,UHTCs_book,Lengauer_jac_1995,Turchanin_sssr_1982,Storms1967refractory,Westrum_jtac_2002} (Fig.~\ref{fig:heat}).
Only small deviations, below 10\% error, are found for TiB$_2$ and HfB$_2$.
It has been pointed out that the origin of this difference is the importance of higher-order lattice anharmonic vibrations at higher temperatures, which are not included under the frame of the quasi-harmonic approximation~\cite{Xiang_jap_2015}.

\noindent \textbf{Gr\"uneisen parameter}. Gr\"uneisen parameter, ${\gamma}$, is a good measurement of the compressibility of the phonons and it is often used to estimate the anharmonicity of  the vibrations in the crystal (Fig.S4).
As other properties already discussed, ${\gamma}$ is a tensorial magnitude which depends on the direction of the tension-compression.
Here, Gr\"uneisen parameter was calculated based on an isotropic expansion-compression of the solid,  
\begin{equation}
 \label{gru}
\bar\gamma = \frac{\sum_{{\bf q}, j}\gamma_{{\bf q},j} c_{{\bf q}, j}}{\sum_{{\bf q}, j}c_{{\bf q},j}},
\end{equation}
where,
\begin{equation}
 \label{gru}
\gamma_{{\bf q},j} = \frac{V^{0K}_{\mathrm{eq}}}{2\omega_{{\bf q},j}^2}\sum_j e_{{\bf q},j}  \frac{\partial D_{\bf q}}{\partial V}e_{{\bf q},j}.
\end{equation}
$D_{\bf q}$ is the dynamical matrix for a wave-vector, ${\bf q}$, $\omega_{{\bf q},j}$ is the vibrational frequency, and $e_{{\bf q},j}$ is the eigenvector for phonon branch, $j$.
The detailed procedure can be found in Refs.~\cite{pinku_prm_2019,curtarolo:art114}.
It seems that the calculated values for borides follow the same trend as experimental results reported by Wiley {\it et al.}~\cite{Wiley_jlcm_1969}, with $\bar\gamma$ being constant over 500~K.
Quantitatively, calculated values seems to slightly overestimated  Wiley {\it et al.} results, however, values reported by Ajami {\it et al.}~\cite{Ajami_jlcm_1974} and Dodd {\it et al.}~\cite{Dodd_jms_2003} at 300~K are in good agreement with the predictions, indicating that the error of the calculation is lower than the deviation of the experimental measurements. 
In addition to the comparison with the experimental results, different conclusions can be extracted from Fig.S4.
While no large changes are observed with the temperature for carbides and nitrides, borides present the higher dependence with respect to the temperature at low temperatures.
For instance, ${\bar\gamma}$ for ZrB$_2$ is reduced up to a 30\% from its maximum value at 80~K to 2000~K.
If families are compared to each other, nitrides present the higher values for ${\bar\gamma}$, then carbides and finally borides.

\begin{figure*}[hbtp!]
  \includegraphics[width=0.95\textwidth]{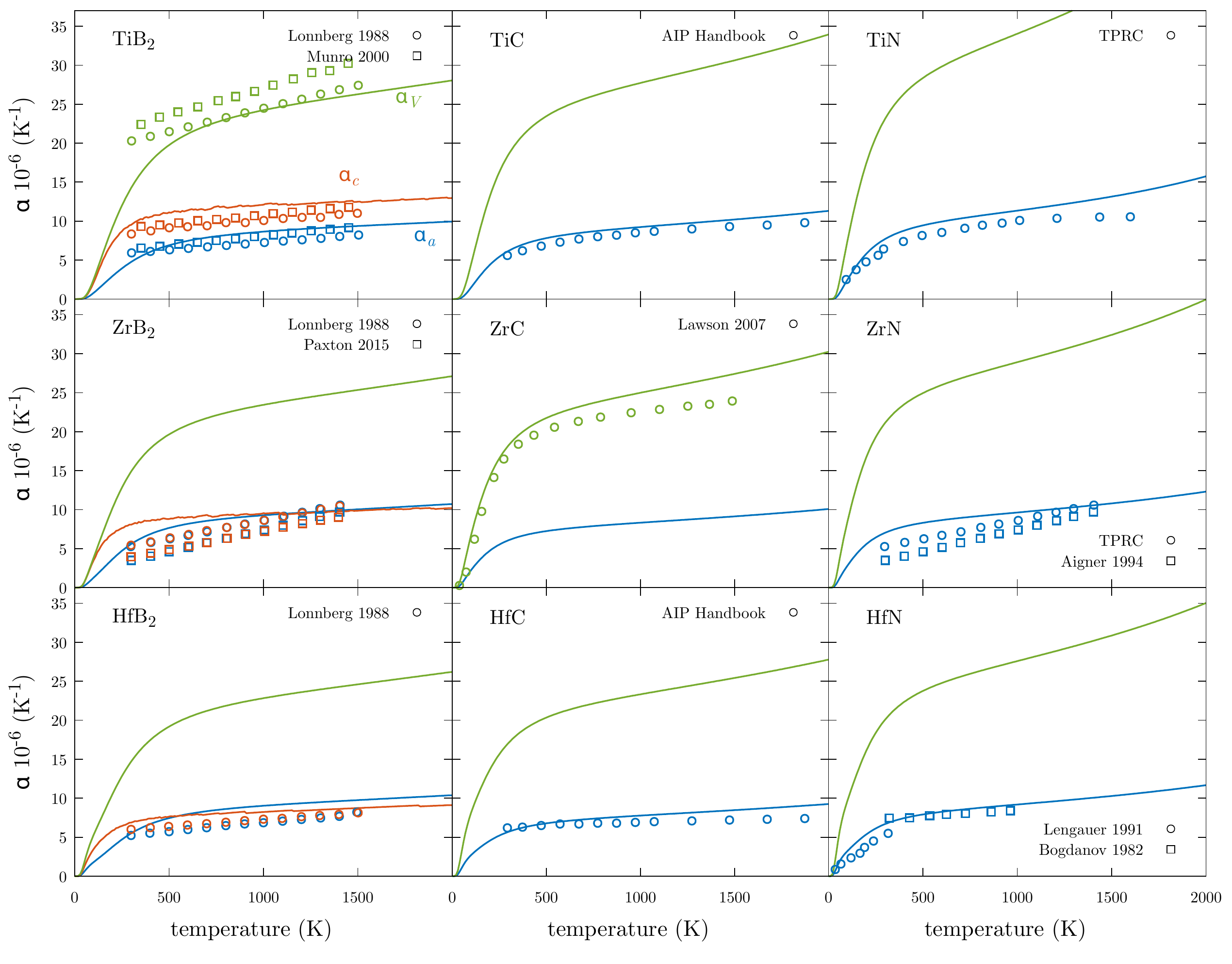}  \vspace{-2mm}
  \caption{\small
   Volumetric thermal expansión coefficient, $\alpha_V$ (green), and linear thermal expansión coefficients, $\alpha_a$ (blue) and $\alpha_c$ (orange) for UHTCs. Solid lines represent calculated values while open points represent experimental measurements.
 }
  \label{fig:alpha}
\end{figure*}

\noindent \textbf{Thermal Expansion}.
Some thermodynamic properties such as heat capacity, can be accurately obtained with low computationally demanding methods such as \GIBBS~\cite{BlancoGIBBS2004,curtarolo:art96}.
However, the accurate prediction of thermal expansion coefficient requires the calculation of the free energy surface which is computationally demanding.
Alternatives to the standard \QHA~\cite{Carrier_prb_2007,Baroni_rmg_2010,curtarolo:art114} such as \QHAPPP~\cite{pinku_prm_2019} can reduce the computational cost, but \QHAPPP\ has been only used for calculating volumetric thermal expansion and cannot capture the anisotropic nature of the material, when isotropic deformations are applied. 
The framework developed in this work fills this gap, capturing the anisotropy of the system through the calculation of linear thermal expansion coefficients and reducing the computational cost to the same level than \QHAPPP.

Linear and volumetric thermal expansion coefficient, $\alpha_i$, are calculated using the free energy curves obtained for the temperature-dependent elastic constants,
\begin{equation}
 \label{alphai}
  \alpha_i = \left( \frac{C_{\epsilon}}{V} \right) \sum^6_{j=1} \left(s_{ij} \right) \gamma_j
\end{equation}
where $C_{\epsilon}$ is the heat capacity at constant strain $\epsilon$, $\gamma_j$ is the Gr\"uneisen parameter along different $j$ directions and $s_{ij}$ are the elastic compliance constants.

When the calculated values are compared with experiments (Fig.~\ref{fig:alpha}), good agreement was obtained for most carbides~\cite{AIP_Handbook,Lawson_pm_2007} and nitrides~\cite{Touloukian_Serie,Aigner_jac_1994,Bogdanov_sov_1982,Lonnberg_jlcm_1988}, while larger deviations were found for borides~\cite{Munro_2000,Lonnberg_jlcm_1988,Paxton_jc_2016}.
As it is defined in Eq.~\ref{alphai}, the $\alpha$ depends on $C_{\epsilon}$, $s_{ij}$ and $\gamma$.
It has been proven in the previous sections that accurate values were obtained for $C_{\epsilon}$ and  $s_{ij}$ ($c_{ij}$). However higher deviations were found for $\gamma$.
Thus, the main source of error stems from the description of the anharmonicity of the material which is not completely well described by the \QHA. 

\section{Conclusions}

In this work, a new framework for the calculation of the thermoelastic properties of materials has been developed.
The main advantage of this new approach relies on the drastic reduction of the computational cost without loosing accuracy.
Ultra-high temperature ceramics have been explored within this framework because of the difficulty of experimentally obtaining their mechanical properties at temperatures close to their melting point. 
Very good agreement between experiments and calculations was found not only for the elastic constants but also for other mechanical properties such as $B$, $G$, $Y$, or $\sigma$.
Although hardness is a property that also depends on plastic deformation, good trends were also predicted.
While approximations or frameworks with similar computational cost only predict isotropic or averaged properties, this new approach also predicts anisotropic properties describing directional mechanical properties such as $Y$ or establishing upper and lower limits for polycrystalline materials.
Thermodynamical and thermal properties were also explored, obtaining, in most cases, good agreement with available experimental data.

This new approach is a perfect tool for the  systematic exploration of the thermoelastic properties of crystalline materials, because of the substantial reduction of the computational cost. 
Moreover, this framework can be considered as a starting point for the development of more elaborated methods in order to solve some deficiencies, such as the inclusion of anharmonic effects or the exploration of more complex systems, such as solid solutions.

\section*{Conflicts of interest}
There are no conflicts to declare.
 
\section*{Acknowledgments}
This work was funded by the Ministerio de Ciencia e Innovación (PID2019-106871GB-I00), and European Union´s Horizon 2020 research and innovation programme under the Marie Sklodowska-Curie grant agreement HT-PHOTO-DB No 752608. 
The authors thankfully acknowledge the computer resources at Lusitania and the technical support provided by Cénits-COMPUTAEX and Red Española de Supercomputación, RES (QS-2019-2-0006, QS-2019-3-0021, QS-2020-2-0033). 

\twocolumngrid
\bibliographystyle{PhysRevwithTitles-noDOI-v1b} 
\bibliography{xpinku-jose} 

\newcommand{\Ozolins}{Ozoli\c{n}\v{s}}
\begin{thebibliography}{100}
\expandafter\ifx\csname urlstyle\endcsname\relax
  \providecommand{\doi}[1]{doi:\discretionary{}{}{}#1}\else
  \providecommand{\doi}{doi:\discretionary{}{}{}\begingroup
  \urlstyle{rm}\Url}\fi
\providecommand{\selectlanguage}[1]{\relax}
\providecommand{\bibAnnoteFile}[1]{%
  \IfFileExists{#1}{\begin{quotation}\noindent\textsc{Key:} #1\\
  \textsc{Annotation:}\ \input{#1}\end{quotation}}{}}
\providecommand{\bibAnnote}[2]{%
  \begin{quotation}\noindent\textsc{Key:} #1\\
  \textsc{Annotation:}\ #2\end{quotation}}

\bibitem{EWuchina_ESI_2007}
E.~Wuchina, E.~Opila, M.~Opeka, W.~Fahrenholtz, and I.~Talmy, \emph{UHTCs:
  Ultra-High Temperature Ceramic materials for extreme environment
  applications}, Electrochemical Society, Inc. \textbf{16}, 30 (2007).
\bibAnnoteFile{EWuchina_ESI_2007}

\bibitem{UHTC_book}
W.~G. Fahrenholtz, E.~J. Wuchina, W.~E. Lee, and Y.~Zhou, \emph{Ultra‐High
  Temperature Ceramics: Materials for extreme environments Applications}
  (Wiley, New York, 2014), 1 edn.
\bibAnnoteFile{UHTC_book}

\bibitem{Fahrenholtz_sm_2017}
W.~G. Fahrenholtz and G.~E. Hilmas, \emph{Ultra-high temperature ceramics:
  Materials for extreme environments}, Scr. Mater. \textbf{129}, 94 (2017).
\bibAnnoteFile{Fahrenholtz_sm_2017}

\bibitem{Moissan_1896}
H.~Moissan, \emph{Sur un nouveau carbure de zirconium}, Comptes rendus
  hebdomadaires des séances de l Académie des Sciences \textbf{122}, 651
  (1896).
\bibAnnoteFile{Moissan_1896}

\bibitem{Tucker_1902}
S.~A. Tucker and H.~R. Moody, \emph{II. - The production of hitherto unknown
  metallic borides}, J. Chem. Soc., Trans. \textbf{81}, 14--17 (1902).
\bibAnnoteFile{Tucker_1902}

\bibitem{Kalish_1969}
D.~Kalish, E.~V. Clougherty, and K.~Kreder, \emph{Strength, Fracture Mode, and
  Thermal Stress Resistance of {HfB$_2$} and {ZrB$_2$}}, J. Am. Ceram. Soc.
  \textbf{52}, 30 (1969).
\bibAnnoteFile{Kalish_1969}

\bibitem{Levine_jecs_2002}
S.~R. Levine, E.~J. Opila, M.~C. Halbig, J.~D. Kiser, M.~Singh, and J.~A.
  Salem, \emph{Evaluation of ultra-high temperature ceramics for aeropropulsion
  use}, J. Eur. Ceram. Soc. \textbf{22}, 2757 (2002).
\bibAnnoteFile{Levine_jecs_2002}

\bibitem{VanWie_jms_2004}
D.~M.~V. Wie, D.~Drewry, D.~E. King, and C.~M. Hudson, \emph{The hypersonic
  environment: Required operating conditions and design challenges}, J. Mater.
  Sci. \textbf{39}, 5915 (2004).
\bibAnnoteFile{VanWie_jms_2004}

\bibitem{Opeka_jecs_1999}
M.~M. Opeka, I.~G. Talmy, E.~J. Wuchina, J.~A. Zaykoski, and S.~J. Causey,
  \emph{Mechanical, Thermal, and Oxidation Properties of Refractory Hafnium and
  zirconium Compounds}, J. Eur. Ceram. Soc. \textbf{19}, 2405 (1999).
\bibAnnoteFile{Opeka_jecs_1999}

\bibitem{Zhang_cms_2008}
X.~Zhang, X.~Luo, J.~Han, J.~L., and W.~Han, \emph{Electronic structure,
  elasticity and hardness of diborides of zirconium and hafnium: First
  principles calculations}, Comput.\ Mater.\ Sci. \textbf{44}, 411 (2008).
\bibAnnoteFile{Zhang_cms_2008}

\bibitem{Chamberlain_jacers_2004}
A.~L. Chamberlain, W.~G. Fahrenholtz, G.~E. Hilmas, and D.~T. Ellerby,
  \emph{High-strength zirconium diboride-based ceramics}, J. Am. Ceram. Soc.
  \textbf{87}, 1170--1172 (2004).
\bibAnnoteFile{Chamberlain_jacers_2004}

\bibitem{Gasch_jms_2004}
M.~Gasch, D.~Ellerby, E.~Irby, S.~Beckman, M.~Gusman, and S.~Johnson,
  \emph{Processing, properties and arc jet oxidation of hafnium
  diboride/silicon carbide ultra high temperature ceramics}, J. Mater. Sci.
  \textbf{39}, 5925--5937 (2004).
\bibAnnoteFile{Gasch_jms_2004}

\bibitem{Tang_msea_2007}
S.~Tang, J.~Deng, S.~Wang, W.~Liu, and K.~Yang, \emph{Ablation behaviors of
  ultra-high temperature ceramic composites}, Mater. Sci. Eng. A \textbf{465},
  1 (2007).
\bibAnnoteFile{Tang_msea_2007}

\bibitem{Monteverde_msea_2008}
F.~Monteverde, A.~Bellosi, and L.~Scatteia, \emph{Processing and properties of
  ultra-high temperature ceramics for space applications}, Mater. Sci. Eng. A
  \textbf{485}, 415 (2008).
\bibAnnoteFile{Monteverde_msea_2008}

\bibitem{Han_cst_2008}
J.~Han, P.~Hu, X.~Zhang, S.~Meng, and W.~Han, \emph{Oxidation-resistant
  {ZrB$_2$-SiC} composites at 2200~\textdegree{}{C}}, Compos. Sci. Technol.
  \textbf{68}, 799 (2008).
\bibAnnoteFile{Han_cst_2008}

\bibitem{Jackson_jacers_2011}
H.~F. Jackson, D.~D. Jayaseelan, D.~Manara, C.~P. Casoni, and W.~E. Lee,
  \emph{Laser Melting of Zirconium Carbide: Determination of Phase Transitions
  in Refractory Ceramic Systems}, J. Am. Ceram. Soc. \textbf{94}, 3561 (2011).
\bibAnnoteFile{Jackson_jacers_2011}

\bibitem{Neuman_jacers_2013}
E.~W. Neuman, G.~E. Hilmas, and W.~G. Fahrenholtz, \emph{Strength of
  {Z}irconium {D}iboride to 2300~\textdegree{}{C}}, J. Am. Ceram. Soc.
  \textbf{96}, 47 (2013).
\bibAnnoteFile{Neuman_jacers_2013}

\bibitem{Miller_jacers_2015}
M.~Miller-Oana, P.~Neff, M.~Valdez, A.~Powell, M.~Packard, L.~S. Walker, and
  E.~L. Corral, \emph{Oxidation Behavior of Aerospace Materials in High
  Enthalpy Flows Using an Oxyacetylene Torch Facility}, J. Am. Ceram. Soc.
  \textbf{98}, 1300 (2015).
\bibAnnoteFile{Miller_jacers_2015}

\bibitem{Paul_aac_2016}
A.~Paul, J.~G.~P. Binner, B.~Vaidhyanathan, A.~C.~J. Heaton, and P.~M. Brown,
  \emph{Heat flux mapping of oxyacetylene flames and their use to characterise
  {Cf-HfB$_2$} composites}, Adv. Appl. Ceram \textbf{115}, 158 (2016).
\bibAnnoteFile{Paul_aac_2016}

\bibitem{Zhao_jssc_2008}
Z.~Erjun and W.~Zhijian, \emph{Electronic and mechanical properties of 5d
  transition metal mononitrides via first principles}, J. Solid State Chem.
  \textbf{181}, 2814 (2008).
\bibAnnoteFile{Zhao_jssc_2008}

\bibitem{Zhang_jac_2011}
J.~D. Zhang, X.~L. Cheng, and D.~H. Li, \emph{First-principles study of the
  elastic and thermodynamic properties of {HfB$_2$} with {AlB$_2$} structure
  under high pressure}, J. Alloy. Compd. \textbf{509}, 9577 (2011).
\bibAnnoteFile{Zhang_jac_2011}

\bibitem{Zeng_prb_2013}
Q.~Zeng, J.~Peng, A.~R. Oganov, Q.~Zhu, C.~Xie, X.~Zhang, D.~Dong, L.~Zhang,
  and L.~Cheng, \emph{Prediction of stable hafnium carbides: Stoichiometries,
  mechanical properties, and electronic structure}, Phys.\ Rev.\ B \textbf{88},
  214107 (2013).
\bibAnnoteFile{Zeng_prb_2013}

\bibitem{Cheng_jacers_2015}
T.~Cheng and W.~Li, \emph{The Temperature-Dependent Ideal Tensile Strength of
  {ZrB$_2$}, {HfB$_2$}, and {TiB$_2$}}, J. Am. Ceram. Soc. \textbf{98}, 190
  (2015).
\bibAnnoteFile{Cheng_jacers_2015}

\bibitem{Xiang_jacers_2017}
H.~Xiang, Z.~Feng, Z.~Li, and Y.~Zhou, \emph{First-principles investigations on
  elevated temperature elastic and thermodynamic properties of {ZrB$_2$} and
  {HfB$_2$}}, J. Am. Ceram. Soc. \textbf{100}, 3662 (2017).
\bibAnnoteFile{Xiang_jacers_2017}

\bibitem{Carrier_prb_2007}
P.~Carrier, R.~Wentzcovitch, and J.~Tsuchiya, \emph{First-principles prediction
  of crystal structures at high temperatures using the quasiharmonic
  approximation}, Phys.\ Rev.\ B \textbf{76}, 064116 (2007).
\bibAnnoteFile{Carrier_prb_2007}

\bibitem{Baroni_rmg_2010}
S.~Baroni, P.~Giannozzi, and E.~Isaev, \emph{Density-Functional Perturbation
  Theory for Quasi-Harmonic Calculations}, Rev. Mineral Geochem. \textbf{71},
  39--57 (2010).
\bibAnnoteFile{Baroni_rmg_2010}

\bibitem{curtarolo:art114}
P.~Nath, J.~J. Plata, D.~Usanmaz, R.~{Al~Rahal~Al~Orabi}, M.~Fornari,
  M.~{Buongiorno Nardelli}, C.~Toher, and S.~Curtarolo, \emph{High-throughput
  prediction of finite-temperature properties using the quasi-harmonic
  approximation}, Comput.\ Mater.\ Sci. \textbf{125}, 82--91 (2016).
\bibAnnoteFile{curtarolo:art114}

\bibitem{Golesorkhtabar_ElaStic_CPC_2013}
R.~Golesorkhtabar, P.~Pavone, J.~Spitaler, P.~Puschnig, and C.~Draxl,
  \emph{ElaStic: A tool for calculating second-order elastic constants from
  first principles}, Comput.\ Phys.\ Commun. \textbf{184}, 1861--1873 (2013).
\bibAnnoteFile{Golesorkhtabar_ElaStic_CPC_2013}

\bibitem{Zhao_prb_2007}
J.~Zhao, J.~M. Winey, and Y.~M. Gupta, \emph{First-principles calculations of
  second- and third-order elastic constants for single crystals of arbitrary
  symmetry}, Phys.\ Rev.\ B \textbf{75}, 094105 (2007).
\bibAnnoteFile{Zhao_prb_2007}

\bibitem{Davies_jpcs_1974}
G.~F. Davies, \emph{Effective elastic moduli under hydro-static stress-I.
  Quasiharmonic theory}, J.\ Phys.\ Chem.\ Solids \textbf{35}, 1513--1520
  (1974).
\bibAnnoteFile{Davies_jpcs_1974}

\bibitem{Karki_science_1999}
B.~B. Karki, R.~M. Wentzcovitch, S.~D. Gironcoli, and S.Baroni,
  \emph{First-principles determination of elastic anisotropy and wave
  velocities of {MgO} at lower mantle conditions}, Science \textbf{286},
  1705--1707 (1999).
\bibAnnoteFile{Karki_science_1999}

\bibitem{Karki_prb_2000}
B.~Karki, R.~Wentzcovitch, S.~de~Gironcoli, and S.~Baroni, \emph{High-pressure
  lattice dynamics and thermoelasticity of {MgO}}, Phys.\ Rev.\ B \textbf{61},
  8793--8800 (2000).
\bibAnnoteFile{Karki_prb_2000}

\bibitem{Wu_prb_2011}
Z.~Wu and R.~M. Wentzcovitch, \emph{Quasiharmonic thermal elasticity of
  crystals: {An} analytical approach}, Phys.\ Rev.\ B \textbf{83} (2011).
\bibAnnoteFile{Wu_prb_2011}

\bibitem{Kadas_prb_2007}
K.~Kádas, L.~Vitos, R.~Ahuja, B.~Johansson, and J.~Kollár,
  \emph{Temperature-dependent elastic properties of $\alpha$-beryllium from
  first principles}, Phys.\ Rev.\ B \textbf{76}, 235109 (2007).
\bibAnnoteFile{Kadas_prb_2007}

\bibitem{Wang_jpcm_2010}
Y.~Wang, J.~J. Wang, H.~Zhang, V.~R. Manga, S.~L. Shang, L.~Q. Chen, and Z.~K.
  Liu, \emph{A first-principles approach to finite temperature elastic
  constants}, J.\ Phys.:\ Condens.\ Matter \textbf{22}, 225404 (2010).
\bibAnnoteFile{Wang_jpcm_2010}

\bibitem{Liu_prb_2019_2}
G.~Liu, Z.~Gao, and J.~Ren, \emph{Anisotropic thermal expansion and
  thermodynamic properties of monolayer $\ensuremath{\beta}$-{T}e}, Phys.\
  Rev.\ B \textbf{99}, 195436 (2019).
\bibAnnoteFile{Liu_prb_2019_2}

\bibitem{Liu_cms_2019}
G.~Liu, H.~Wang, G.~L. Li, and D.~Wang, \emph{Giant anisotropy of thermal
  expansion and thermomechanical properties of monolayer $\alpha$-antimonene: A
  first-principles study}, Comput.\ Mater.\ Sci. \textbf{169}, 109132 (2019).
\bibAnnoteFile{Liu_cms_2019}

\bibitem{Destefanis_m_2019}
M.~Destefanis, C.~Ravoux, A.~Cossard, and A.~Erba, \emph{Thermo-Elasticity of
  Materials from Quasi-Harmonic Calculations}, Minerals \textbf{9}, 16 (2019).
\bibAnnoteFile{Destefanis_m_2019}

\bibitem{ThermoCrys}
D.~C. Wallace, \emph{Thermodynamics of crystals} (Wiley, 1972).
\bibAnnoteFile{ThermoCrys}

\bibitem{SHZhang_cpc_2017}
S.~H. Zhang and R.~F. Zhang, \emph{AELAS: Automatic ELAStic property
  derivations via high-throughput first-principles computation}, Comput.\
  Phys.\ Commun. \textbf{220}, 403 (2017).
\bibAnnoteFile{SHZhang_cpc_2017}

\bibitem{GLUI_prb_2019}
G.~Liu, Z.~Gao, and J.~Ren, \emph{Anisotropic thermal expansion and
  thermodynamic properties of monolayer {$\beta$-Te}}, Phys.\ Rev.\ B
  \textbf{99}, 195436 (2019).
\bibAnnoteFile{GLUI_prb_2019}

\bibitem{Nielsen_prl_1983}
O.~H. Nielsen and R.~M. Martin, \emph{First-Principles Calculation of Stress},
  Phys.\ Rev.\ Lett. \textbf{50}, 697--700 (1983).
\bibAnnoteFile{Nielsen_prl_1983}

\bibitem{curtarolo:art100}
M.~{de~Jong}, W.~Chen, T.~Angsten, A.~Jain, R.~Notestine, A.~Gamst, M.~Sluiter,
  C.~K. Ande, S.~{van~der~Zwaag}, J.~J. Plata, C.~Toher, S.~Curtarolo,
  G.~Ceder, K.~A. Persson, and M.~D. Asta, \emph{Charting the Complete Elastic
  properties of Inorganic Crystalline Compounds}, Sci.\ Data \textbf{2}, 150009
  (2015).
\bibAnnoteFile{curtarolo:art100}

\bibitem{Liu_Cambridge_2016}
Z.~K. Liu and Y.~Wang, \emph{Computational Thermodynamics of Materials}
  (Cambridge University Press, 2016), 1 edn.
\bibAnnoteFile{Liu_Cambridge_2016}

\bibitem{Wang_ACTAMAT_2004}
Y.~Wang, Z.-K. Liu, and L.-Q. Chen, \emph{Thermodynamic properties of {Al},
  {Ni}, {Ni}{Al}, and {Ni}$_{3}${Al} from first-principles calculations}, Acta\
  Mater. \textbf{52}, 2665--2671 (2004).
\bibAnnoteFile{Wang_ACTAMAT_2004}

\bibitem{Duong_jap_2011}
T.~Duong, S.~Gibbons, R.~Kinra, and R.~Arr\'{o}yave, \emph{{\it Ab-initio}
  approach to the electronic, structural, elastic, and finite-temperature
  thermodynamic properties of {Ti}$_{2}{A}{X}$ ({$A$} = {Al} or {Ga} and {$X$}
  = {C} or {N})}, J.\ Appl.\ Phys. \textbf{110}, 093504 (2011).
\bibAnnoteFile{Duong_jap_2011}

\bibitem{vasp_prb1996}
G.~Kresse and J.~Furthm\"{u}ller, \emph{Efficient iterative schemes for {\it ab
  initio} total-energy calculations using a plane-wave basis set}, Phys.\ Rev.\
  B \textbf{54}, 11169--11186 (1996).
\bibAnnoteFile{vasp_prb1996}

\bibitem{quantum_espresso_2009}
P.~Giannozzi, S.~Baroni, N.~Bonini, M.~Calandra, R.~Car, C.~Cavazzoni,
  D.~Ceresoli, G.~L. Chiarotti, M.~Cococcioni, I.~Dabo, A.~{Dal Corso}, S.~{de
  Gironcoli}, S.~Fabris, G.~Fratesi, R.~Gebauer, U.~Gerstmann, C.~Gougoussis,
  A.~Kokalj, M.~Lazzeri, L.~Martin-Samos, N.~Marzari, F.~Mauri, R.~Mazzarello,
  S.~Paolini, A.~Pasquarello, L.~Paulatto, C.~Sbraccia, S.~Scandolo,
  G.~Sclauzero, A.~P. Seitsonen, A.~Smogunov, P.~Umari, and R.~M. Wentzcovitch,
  \emph{{QUANTUM ESPRESSO}: a modular and open-source software project for
  quantum simulations of materials}, J.\ Phys.:\ Condens.\ Matter \textbf{21},
  395502 (2009).
\bibAnnoteFile{quantum_espresso_2009}

\bibitem{pinku_prm_2019}
P.~Nath, D.~Usanmaz, D.~Hicks, C.~Oses, M.~Fornari, M.~N. Buongiorno, C.~Toher,
  and S.~Curtarolo, \emph{{AFLOW-QHA3P}/: Robust and automated method to
  compute thermodynamic properties of solids}, Phys. Rev. Materials \textbf{3},
  073801 (2019).
\bibAnnoteFile{pinku_prm_2019}

\bibitem{Togo_scrmat_2015}
A.~Togo and I.~Tanaka, \emph{First principles phonon calculations in materials
  science}, Scr.\ Mater. \textbf{108}, 1--5 (2015).
\bibAnnoteFile{Togo_scrmat_2015}

\bibitem{Arroyave_ACTAMAT_2005}
R.~Arroyave, D.~Shin, and Z.-K. Liu, \emph{{\it Ab initio} thermodynamic
  properties of stoichiometric phases in the {Ni}-{Al} system}, Acta\ Mater.
  \textbf{53}, 1809--1819 (2005).
\bibAnnoteFile{Arroyave_ACTAMAT_2005}

\bibitem{Huang_cms_2016}
L.~F. Huang, X.~Z. Lu, E.~Tennessen, and J.~M. Rondinelli, \emph{An efficient
  {\it ab-initio} quasiharmonic approach for the thermodynamics of solids},
  Comput.\ Mater.\ Sci. \textbf{120}, 84--93 (2016).
\bibAnnoteFile{Huang_cms_2016}

\bibitem{Featherston_pr_1963}
F.~H. Featherston and J.~R. Neighbours, \emph{Elastic Constants of Tantalum,
  Tungsten, and Molybdenum}, Phys.\ Rev. \textbf{130}, 1324--1333 (1963).
\bibAnnoteFile{Featherston_pr_1963}

\bibitem{Davenport_prb_1999}
T.~Davenport, L.~Zhou, and J.~Trivisonno, \emph{Ultrasonic and atomic force
  studies of the martensitic transformation induced by temperature and uniaxial
  stress in NiAl alloys}, Phys.\ Rev.\ B \textbf{59}, 3421--3426 (1999).
\bibAnnoteFile{Davenport_prb_1999}

\bibitem{Basaadat_ijmpc_2020}
M.~Basaadat and M.~Payami, \emph{Elastic stiffness tensors of {Zr–$x$Nb}
  alloy in the presence of defects: A molecular dynamics study}, Int. J. Mod.
  Phys. C \textbf{0}, 2050028 (0).
\bibAnnoteFile{Basaadat_ijmpc_2020}

\bibitem{spglib_python}
\emph{Spglib for Python}, https://atztogo.github.io/spglib/python-spglib.html
  (2017).
\bibAnnoteFile{spglib_python}

\bibitem{kresse_vasp}
G.~Kresse and J.~Hafner, \emph{{\it Ab initio} molecular dynamics for liquid
  metals}, Phys.\ Rev.\ B \textbf{47}, 558--561 (1993).
\bibAnnoteFile{kresse_vasp}

\bibitem{PAW}
P.~E. Bl\"{o}chl, \emph{Projector augmented-wave method}, Phys.\ Rev.\ B
  \textbf{50}, 17953--17979 (1994).
\bibAnnoteFile{PAW}

\bibitem{PBE}
J.~P. Perdew, K.~Burke, and M.~Ernzerhof, \emph{Generalized Gradient
  Approximation Made Simple}, Phys.\ Rev.\ Lett. \textbf{77}, 3865--3868
  (1996).
\bibAnnoteFile{PBE}

\bibitem{Calderon_cms_2015}
C.~E. Calderon, J.~J. Plata, C.~Toher, C.~Oses, O.~Levy, M.~Fornari, A.~Natan,
  M.~J. Mehl, G.~L.~W. Hart, M.~{Buongiorno Nardelli}, and S.~Curtarolo,
  \emph{The {AFLOW} standard for high-throughput materials science
  calculations}, Comput.\ Mater.\ Sci. \textbf{108 Part A}, 233--238 (2015).
\bibAnnoteFile{Calderon_cms_2015}

\bibitem{Tadano_jpcm_2014}
T.~Tadano, Y.~Gohda, and S.~Tsuneyuki, \emph{Anharmonic force constants
  extracted from first-principles molecular dynamics: applications to heat
  transfer simulations}, J.\ Phys.:\ Condens.\ Matter \textbf{26}, 225402
  (2014).
\bibAnnoteFile{Tadano_jpcm_2014}

\bibitem{curtarolo:art125}
J.~J. Plata, P.~Nath, D.~Usanmaz, J.~Carrete, C.~Toher, M.~{de Jong}, M.~D.
  Asta, M.~Fornari, M.~{Buongiorno Nardelli}, and S.~Curtarolo, \emph{An
  efficient and accurate framework for calculating lattice thermal conductivity
  of solids: {AFLOW}-{AAPL} {Au}tomatic {A}nharmonic {P}honon {Li}brary}, NPJ\
  Comput.\ Mater. \textbf{3}, 45 (2017).
\bibAnnoteFile{curtarolo:art125}

\bibitem{Alfe_cpc_2009}
D.~Alf\'{e}, \emph{{PHON}: A program to calculate phonons using the small
  displacement method}, Comput.\ Phys.\ Commun. \textbf{180}, 2622--2633
  (2009).
\bibAnnoteFile{Alfe_cpc_2009}

\bibitem{Isaev_jap_2007}
E.~I. Isaev, S.~I. Simak, I.~A. Abrikosov, R.~Ahuja, Y.~K. Vekilov, M.~I.
  Katsnelson, A.~I. Lichtenstein, and B.~Johansson, \emph{Phonon related
  properties of transition metals, their carbides, and nitrides: {A}
  first-principles study}, J.\ Appl.\ Phys. \textbf{101}, 123519 (2007).
\bibAnnoteFile{Isaev_jap_2007}

\bibitem{Aizawa_prb_2001}
T.~Aizawa, W.~Hayami, and S.~Otani, \emph{Surface phonon dispersion of
  ${\mathrm{ZrB}}_{2}(0001)$ and ${\mathrm{NbB}}_{2}(0001)$}, Phys.\ Rev.\ B
  \textbf{65}, 024303 (2001).
\bibAnnoteFile{Aizawa_prb_2001}

\bibitem{Pintschovius_jpcss_1978}
L.~Pintschovius, W.~Reichardt, and B.~Scheerer, \emph{Lattice dynamics of
  {TiC}}, J.\ Phys.\ C:\ Solid\ State\ Phys. \textbf{11}, 1557--1562 (1978).
\bibAnnoteFile{Pintschovius_jpcss_1978}

\bibitem{Kress_prb_1978}
W.~Kress, P.~Roedhammer, H.~Bilz, W.~D. Teuchert, and A.~N. Christensen,
  \emph{Phonon anomalies in transition-metal nitrides: {TiN}}, Phys.\ Rev.\ B
  \textbf{17}, 111--113 (1978).
\bibAnnoteFile{Kress_prb_1978}

\bibitem{Weber_prb_1973}
W.~Weber, \emph{Lattice Dynamics of Transition-Metal Carbides}, Phys.\ Rev.\ B
  \textbf{8}, 5082--5092 (1973).
\bibAnnoteFile{Weber_prb_1973}

\bibitem{Christensen_prb_1979}
A.~N. Christensen, O.~W. Dietrich, W.~Kress, and W.~D. Teuchert, \emph{Phonon
  anomalies in transition-metal nitrides: {ZrN}}, Phys.\ Rev.\ B \textbf{19},
  5699--5703 (1979).
\bibAnnoteFile{Christensen_prb_1979}

\bibitem{Christensen_prb_1983}
A.~N. Christensen, W.~Kress, M.~Miura, and N.~Lehner, \emph{Phonon anomalies in
  transition-metal nitrides: HfN}, Phys.\ Rev.\ B \textbf{28}, 977--981 (1983).
\bibAnnoteFile{Christensen_prb_1983}

\bibitem{Smith_proc_1971}
H.~G. Smith and W.~Glazer, in \emph{Proceedings of the International Conference
  on Phonons}, edited by M.~A. Nusimovici (Wiley, Rennes, Frances, 1971).
\bibAnnoteFile{Smith_proc_1971}

\bibitem{Smith_prl_1972}
H.~G. Smith, \emph{Phonon Anomalies in Transition-Metal Carbides}, Phys.\ Rev.\
  Lett. \textbf{29}, 353--354 (1972).
\bibAnnoteFile{Smith_prl_1972}

\bibitem{Okamoto_mrssp_2003}
N.~L. Okamoto, M.~Kusakari, K.~Tanaka, H.~Inui, M.~Yamaguchi, and S.~Otani,
  \emph{Mechanical and thermal properties of single crystals of {ZrB$_2$}}
  (2003), vol. 753, pp. 83--88.
\bibAnnoteFile{Okamoto_mrssp_2003}

\bibitem{Okamoto_am_2010}
N.~L. Okamoto, M.~Kusakari, K.~Tanaka, H.~Inui, and S.~Otani, \emph{Anisotropic
  elastic constants and thermal expansivities in monocrystal {CrB$_2$},
  {TiB$_2$}, and {ZrB$_2$}}, Acta\ Mater. \textbf{58}, 76 -- 84 (2010).
\bibAnnoteFile{Okamoto_am_2010}

\bibitem{Xiang_jap_2015}
H.~Xiang, Z.~Feng, Z.~Li, and Y.~Zhou, \emph{Temperature-dependence of
  structural and mechanical properties of {TiB$_2$}: A first principle
  investigation}, J.\ Appl.\ Phys. \textbf{117}, 225902 (2015).
\bibAnnoteFile{Xiang_jap_2015}

\bibitem{Chang_jap_1966}
R.~Chang and L.~J. Graham, \emph{Low‐Temperature Elastic Properties of {ZrC}
  and {TiC}}, J.\ Appl.\ Phys. \textbf{37}, 3778--3783 (1966).
\bibAnnoteFile{Chang_jap_1966}

\bibitem{Chen_pnas_2005}
X.~J. Chen, V.~V. Struzhkin, Z.~Wu, M.~Somayazulu, J.~Qian, S.~Kung, A.~N.
  Christensen, Y.~Zhao, R.~E. Cohen, H.~Mao, and R.~J. Hemley, \emph{Hard
  superconducting nitrides}, Proc.\ Natl.\ Acad.\ Sci. \textbf{102}, 3198--3201
  (2005).
\bibAnnoteFile{Chen_pnas_2005}

\bibitem{Holec_prb_2012}
D.~Holec, M.~Fri\'ak, J.~Neugebauer, and P.~H. Mayrhofer, \emph{Trends in the
  elastic response of binary early transition metal nitrides}, Phys.\ Rev.\ B
  \textbf{85}, 064101 (2012).
\bibAnnoteFile{Holec_prb_2012}

\bibitem{Steneteg_prb_2015}
P.~Steneteg, O.~Hellman, O.~Y. Vekilova, N.~Shulumba, F.~Tasn\'adi, and
  A.~Abrikosov, \emph{Temperature dependence of {TiN} elastic constants from ab
  initio molecular dynamics simulations}, Phys.\ Rev.\ B \textbf{87}, 094114
  (2013).
\bibAnnoteFile{Steneteg_prb_2015}

\bibitem{zhang_arXiv_2020}
J.~Zhang and J.~M. McMahon, \emph{Temperature-dependent mechanical properties
  of {ZrC} and {HfC} from first principles} (2020).
\bibAnnoteFile{zhang_arXiv_2020}

\bibitem{Karki_jpcm_1997}
B.~B. Karki, G.~J. Ackland, and J.~Crain, \emph{Elastic instabilities in
  crystals from ab initio stress–strain relations}, J.\ Phys.:\ Condens.\
  Matter \textbf{9}, 8579--8589 (1997).
\bibAnnoteFile{Karki_jpcm_1997}

\bibitem{Born_Dynamic}
M.~Born and K.~Huang, \emph{Dynamic Theory of Crystal Lattices} (Oxfird
  University Press., Oxford, 1954).
\bibAnnoteFile{Born_Dynamic}

\bibitem{Wang_prl_1993}
J.~Wang, S.~Yip, S.~R. Phillpot, and D.~Wolf, \emph{Crystal instabilities at
  finite strain}, Phys.\ Rev.\ Lett. \textbf{71}, 4182 (1993).
\bibAnnoteFile{Wang_prl_1993}

\bibitem{Nye_Crystals}
J.~F. Nye, \emph{Physical Properties of Crystals: Their Representation by
  Tensors and Matrices} (Oxfird University Press., Oxford, 1985).
\bibAnnoteFile{Nye_Crystals}

\bibitem{Mouhat_prb_2014}
F.~Mouhat and F.~X. Coudert, \emph{Necessary and sufficient elastic stability
  conditions in various crystal systems}, Phys.\ Rev.\ B \textbf{90}, 224104
  (2014).
\bibAnnoteFile{Mouhat_prb_2014}

\bibitem{Voigt_book}
W.~Voigt, \emph{Lehrbuch der Kristallphysik} (Teubner, Leipsig, 1928).
\bibAnnoteFile{Voigt_book}

\bibitem{Reuss_ZAMM_1929}
A.~Reuss, \emph{Berechnung der Fließgrenze von Mischkristallen auf Grund der
  Plastizitätsbedingung für Einkristalle}, Z. Angew. Math. Mech. \textbf{9},
  49--58 (1929).
\bibAnnoteFile{Reuss_ZAMM_1929}

\bibitem{Hill_pps_1952}
R.~Hill, \emph{The elastic behaviour of a crystalline aggregate}, Proc. Phys.
  Soc. \textbf{65}, 349--354 (1952).
\bibAnnoteFile{Hill_pps_1952}

\bibitem{Spriggs_jacers_1961}
R.~M. Spriggs, \emph{Expression for Effect of Porosity on Elastic Modulus of
  Polycrystalline Refractory Materials, Particularly Aluminum Oxide}, J.\
  Amer.\ Ceram.\ Soc. \textbf{44}, 628--629 (1961).
\bibAnnoteFile{Spriggs_jacers_1961}

\bibitem{Nanjangud_jacers_1995}
S.~C. Nanjangud, R.~Brezny, and D.~J. Green, \emph{Strength and {Y}oung's
  Modulus Behavior of a Partially Sintered Porous Alumina}, J.\ Amer.\ Ceram.\
  Soc. \textbf{78}, 266--268 (1995).
\bibAnnoteFile{Nanjangud_jacers_1995}

\bibitem{Rice_kem_1996}
R.~W. Rice, \emph{The Porosity Dependence of Physical Properties of Materials:
  A Summary Review}, Key Eng. Mat. \textbf{115}, 1--20 (1996).
\bibAnnoteFile{Rice_kem_1996}

\bibitem{Gibson_prsla_1982}
I.~J. Gibson and M.~F. Ashby, \emph{The mechanics of three-dimensional cellular
  materials}, Proc. R. Soc. Lond. A. Math. Phy. \textbf{382}, 43--59 (1982).
\bibAnnoteFile{Gibson_prsla_1982}

\bibitem{Wiley_jlcm_1969}
D.~E. Wiley, W.~R. Manning, and O.~Hunter, \emph{Elastic properties of
  polycrystalline {TiB$_2$, ZrBi$_2$} and {HfB$_2$} from room temperature to
  1300~{K}}, J.\ Less-Common\ Met. \textbf{18}, 149 -- 157 (1969).
\bibAnnoteFile{Wiley_jlcm_1969}

\bibitem{Pan_jacers_1997}
M.~J. Pan, P.~A. Hoffman, D.~J. Green, and J.~R. Hellmann, \emph{Elastic
  Properties and Microcracking Behavior of Particulate Titanium
  Diboride–Silicon Carbide Composites}, J.\ Amer.\ Ceram.\ Soc. \textbf{80},
  692--698 (1997).
\bibAnnoteFile{Pan_jacers_1997}

\bibitem{Dodd_jms_2003}
S.~P. Dodd, M.~Cankurtaran, and B.~James, \emph{Ultrasonic determination of the
  elastic and nonlinear acoustic properties of transition-metal carbide
  ceramics: {TiC} and {TaC}}, J.\ Mater.\ Sci. \textbf{38}, 1107--1115 (2003).
\bibAnnoteFile{Dodd_jms_2003}

\bibitem{Chen_amse_2009}
D.~Chen, J.~Chen, Y.~Zhao, B.~Yu, C.~Wang, and D.~Shi, \emph{Theoretical study
  of the elastic properties of titanium nitride}, Acta Metall. Sin.-Engl.
  \textbf{22}, 146 -- 152 (2009).
\bibAnnoteFile{Chen_amse_2009}

\bibitem{Mohammadpour_ms_2018}
E.~Mohammadpour, M.~Altarawneh, J.~Al-Nu’airat, Z.~T. Jiang, N.~Mondinos, and
  B.~Z. Dlugogorski, \emph{Thermo-mechanical properties of cubic titanium
  nitride}, Mol. Simulat. \textbf{44}, 415--423 (2018).
\bibAnnoteFile{Mohammadpour_ms_2018}

\bibitem{Kim_jac_2012}
J.~Kim and S.~Kang, \emph{First principles investigation of temperature and
  pressure dependent elastic properties of {ZrC} and {ZrN} using
  {Debye-Gruneisen} theory}, J.\ Alloys\ Compd. \textbf{540}, 94 -- 99 (2012).
\bibAnnoteFile{Kim_jac_2012}

\bibitem{Spoor_apl_1997}
P.~S. Spoor, J.~D. Maynard, M.~J. Pan, D.~J. Green, J.~R. Hellmann, and
  T.~Tanaka, \emph{Elastic constants and crystal anisotropy of titanium
  diboride}, Appl.\ Phys.\ Lett. \textbf{70}, 1959--1961 (1997).
\bibAnnoteFile{Spoor_apl_1997}

\bibitem{Munro_2000}
R.~G. Munro, \emph{Material Properties of Titanium Diboride}, J. Res. Natl.
  Inst. Stand. Technol. \textbf{105}, 709--720 (2000).
\bibAnnoteFile{Munro_2000}

\bibitem{Baranov_sm_1973}
V.~M. Baranov, V.~I. Knyazev, and O.~S. Korostin, \emph{The temperature
  dependence of the elastic constants of nonstoichiometric zirconium carbides},
  Strength Mater. \textbf{5}, 1074--1077 (1973).
\bibAnnoteFile{Baranov_sm_1973}

\bibitem{Travushkin_sm_1973}
G.~G. Travushkin, V.~I. Knyazev, V.~S. Belov, and G.~A. Rymashevskii,
  \emph{Temperature threshold of brittle failure in interstitial phases},
  Strength Mater. \textbf{5}, 639--641 (1973).
\bibAnnoteFile{Travushkin_sm_1973}

\bibitem{Ljungcrantz_jap_1996}
H.~Ljungcrantz, M.~Odén, L.~Hultman, J.~E. Greene, and J.~E. Sundgren,
  \emph{Nanoindentation studies of single‐crystal (001)‐, (011)‐, and
  (111)‐oriented {TiN} layers on {MgO}}, J.\ Appl.\ Phys. \textbf{80},
  6725--6733 (1996).
\bibAnnoteFile{Ljungcrantz_jap_1996}

\bibitem{Kral_Lengauer_1998}
C.~Kral, W.~Lengauer, D.~Rafaja, and P.~Ettmayer, \emph{Critical review on the
  elastic properties of transition metal carbides, nitrides and carbonitrides},
  J.\ Alloys\ Compd. \textbf{265}, 215--233 (1998).
\bibAnnoteFile{Kral_Lengauer_1998}

\bibitem{Yang_jac_2000}
Q.~Yang, W.~Lengauer, T.~Koch, M.~Scheerer, and I.~Smid, \emph{Hardness and
  elastic properties of {Ti(C$_x$N$_{1-x}$), Zr(C$_x$N$_{1-x}$) and
  Hf(C$_x$N$_{1-x}$)}}, J.\ Alloys\ Compd. \textbf{309}, L5--L9 (2000).
\bibAnnoteFile{Yang_jac_2000}

\bibitem{Yu_ijrmh_1999}
Y.~V. Milman, S.~I. Chugunova, I.~V. Goncharova, T.~Chudoba, W.~Lojkowski, and
  W.~Gooch, \emph{Temperature dependence of hardness in silicon–carbide
  ceramics with different porosity}, Int. J. Refract. Met. H. \textbf{17}, 361
  -- 368 (1999).
\bibAnnoteFile{Yu_ijrmh_1999}

\bibitem{Rice_jacers_1994}
R.~W. Rice, C.~C. Wu, and F.~Boichelt, \emph{Hardness–Grain-Size Relations in
  Ceramics}, J.\ Amer.\ Ceram.\ Soc. \textbf{77}, 2539--2553 (1994).
\bibAnnoteFile{Rice_jacers_1994}

\bibitem{Gong_msea_2001}
J.~Gong, H.~Miao, Z.~Zhao, and Z.~Guan, \emph{Load-dependence of the measured
  hardness of {Ti(C,N)}-based cermets}, Mat.\ Sci.\ Eng.\ A \textbf{303}, 179
  -- 186 (2001).
\bibAnnoteFile{Gong_msea_2001}

\bibitem{Guo_jmpt_2018}
B.~Guo, L.~Zhang, L.~Cao, T.~Zhang, F.~Jiang, and L.~Yan, \emph{The correction
  of temperature-dependent Vickers hardness of cemented carbide base on the
  developed high-temperature hardness tester}, J. Mater. Process. Technol.
  \textbf{255}, 426 -- 433 (2018).
\bibAnnoteFile{Guo_jmpt_2018}

\bibitem{Gao_prl_2003}
F.~Gao, J.~He, E.~Wu, S.~Liu, D.~Yu, D.~Li, S.~Zhang, and Y.~Tian,
  \emph{Hardness of Covalent Crystals}, Phys.\ Rev.\ Lett. \textbf{91}, 015502
  (2003).
\bibAnnoteFile{Gao_prl_2003}

\bibitem{Simunek_prl_2006}
A.~{\u S}im{\r u}nek and J.~Vack{\' a}{\u r}, \emph{Hardness of Covalent and
  Ionic Crystals: First-Principle Calculations}, Phys.\ Rev.\ Lett.
  \textbf{96}, 085501 (2006).
\bibAnnoteFile{Simunek_prl_2006}

\bibitem{Chen_intermetallics_2011}
X.~Q. Chen, H.~Niu, D.~Li, and Y.~Li, \emph{Modeling hardness of
  polycrystalline materials and bulk metallic glasses}, Intermetallics
  \textbf{19}, 1275--1281 (2011).
\bibAnnoteFile{Chen_intermetallics_2011}

\bibitem{Chen_prb_2011}
X.-Q. Chen, H.~Niu, C.~Franchini, D.~Li, and Y.~Li, \emph{Hardness of
  $T$-carbon: Density functional theory calculations}, Phys.\ Rev.\ B
  \textbf{84}, 121405 (2011).
\bibAnnoteFile{Chen_prb_2011}

\bibitem{Tian_refmat_2012}
Y.~Tian, B.~Xu, and Z.~Zhao, \emph{Microscopic theory of hardness and design of
  novel superhard crystals}, Int. J. Refract. Met. Hard Mater. \textbf{33},
  93--106 (2012).
\bibAnnoteFile{Tian_refmat_2012}

\bibitem{Shackelford_CRC}
J.~F. Shackelford, Y.~H. Han, S.~Kim, and S.~H. Kwon, \emph{CRC Materials
  Science and Engineering Handbook} (CRC Press, New York, 2015).
\bibAnnoteFile{Shackelford_CRC}

\bibitem{Otani_jcsj_2000}
S.~Otani, \emph{High Temperature Hardness and Flux Growth of {TiB$_2$, VB$_2$}
  and {CrB$_2$} Crystals}, J. Ceram. Soc. Jpn. \textbf{108}, 955--956 (2000).
\bibAnnoteFile{Otani_jcsj_2000}

\bibitem{Raju_jecs_2009}
G.~B. Raju, B.~Basu, N.~H. Tak, and S.~J. Cho, \emph{Temperature dependent
  hardness and strength properties of {TiB$_2$} with {TiSi$_2$} sinter-aid}, J.
  Eur. Ceram. Soc. \textbf{29}, 2119 -- 2128 (2009).
\bibAnnoteFile{Raju_jecs_2009}

\bibitem{Fahrenholtz_jacers_2007}
W.~G. Fahrenholtz, G.~E. Hilmas, I.~G. Talmy, and J.~A. Zaykoski,
  \emph{Refractory Diborides of Zirconium and Hafnium}, J.\ Amer.\ Ceram.\ Soc.
  \textbf{90}, 1347--1364 (2007).
\bibAnnoteFile{Fahrenholtz_jacers_2007}

\bibitem{Xuan_jpd_2002}
Y.~Xuan, C.~H. Chen, and S.~Otani, \emph{High temperature microhardness of
  {ZrB$_2$} single crystals}, J.\ Phys.\ D:\ Appl.\ Phys. \textbf{35},
  L98--L100 (2002).
\bibAnnoteFile{Xuan_jpd_2002}

\bibitem{Csanadi_jecs_2016}
T.~Csanádi, S.~Grasso, A.~Kovalčíková, J.~Dusza, and M.~Reece,
  \emph{Nanohardness and elastic anisotropy of {ZrB$_2$} crystals}, J. Eur.
  Ceram. Soc. \textbf{36}, 239 -- 242 (2016).
\bibAnnoteFile{Csanadi_jecs_2016}

\bibitem{Bsenko_jlcm_1974}
L.~Bsenko and T.~Lundström, \emph{The high-temperature hardness of {ZrB$_2$
  and HfB$_2$}}, J.\ Less-Common\ Met. \textbf{34}, 273 -- 278 (1974).
\bibAnnoteFile{Bsenko_jlcm_1974}

\bibitem{Liang_ci_2019}
H.~Liang, S.~Guan, X.~Li, A.~Liang, Y.~Zeng, C.~Liu, H.~Chen, W.~Lin, D.~He,
  L.~Wang, and F.~Peng, \emph{Microstructure evolution, densification behavior
  and mechanical properties of nano-{HfB$_2$} sintered under high pressure},
  Ceram. Int. \textbf{45}, 7885 -- 7893 (2019).
\bibAnnoteFile{Liang_ci_2019}

\bibitem{Kumashiro_jms_1977}
Y.~Kumashiro, A.~Itoh, T.~Kinoshita, and M.~Sobajima, \emph{The micro-Vickers
  hardness of {TiC} single crystals up to 1500~\textdegree{}{C}}, J. Mater.
  Sci. \textbf{12}, 595--601 (1977).
\bibAnnoteFile{Kumashiro_jms_1977}

\bibitem{Maerky_msea_1996}
C.~Maerky, M.~O. Guillou, J.~L. Henshall, and R.~M. Hooper, \emph{Indentation
  hardness and fracture toughness in single crystal {TiC$_{0.96}$}}, Mat.\
  Sci.\ Eng.\ A \textbf{209}, 329 -- 336 (1996).
\bibAnnoteFile{Maerky_msea_1996}

\bibitem{Kohlstedt_jms_1973}
D.~L. Kohlstedt, \emph{The temperature dependence of microhardness of the
  transition-metal carbides}, J.\ Mater.\ Sci. \textbf{8}, 777 -- 786 (1973).
\bibAnnoteFile{Kohlstedt_jms_1973}

\bibitem{Kumashiro_jmsl_1982}
Y.~Kumashiro, Y.~Nagai, and H.~Kato, \emph{The Vickers microhardness of {NbC},
  {ZrC} and {TaC} single crystals up to 1500}, J. Mater. Sci. Lett. \textbf{1},
  49 -- 52 (1982).
\bibAnnoteFile{Kumashiro_jmsl_1982}

\bibitem{Balko_jecs_2017}
J.~Balko, T.~Csanádi, R.~Sedlák, M.~Vojtko, A.~KovalLíková, K.~Koval,
  P.~Wyzga, and A.~N. Duszová, \emph{Nanoindentation and tribology of {VC},
  {NbC} and {ZrC} refractory carbides}, J. Eur. Ceram. Soc. \textbf{37}, 4371
  -- 4377 (2017).
\bibAnnoteFile{Balko_jecs_2017}

\bibitem{Cheng_jecs_2017}
E.~J. Cheng, Y.~Li, J.~Sakamoto, S.~Han, H.~Sun, J.~Noble, H.~Katsui, and
  T.~Goto, \emph{Mechanical properties of individual phases of {ZrB$_2$-ZrC}
  eutectic composite measured by nanoindentation}, J. Eur. Ceram. Soc.
  \textbf{37}, 4223 -- 4227 (2017).
\bibAnnoteFile{Cheng_jecs_2017}

\bibitem{ASM_book}
M.~Bauccio, ed., \emph{{ASM} Engineered Materials Reference Book} (ASM
  International, Materials Park, Ohio, USA, 1994).
\bibAnnoteFile{ASM_book}

\bibitem{Andrievski_nanom_1997}
R.~A. Andrievski, \emph{Physical-mechanical properties of nanostructured
  titanium nitride}, Nanostruc. Mater. \textbf{9}, 607 -- 610 (1997).
\bibAnnoteFile{Andrievski_nanom_1997}

\bibitem{Mei_jvsta_2013}
A.~B. Mei, B.~M. Howe, C.~Zhang, M.~Sardela, J.~N. Eckstein, L.~Hultman,
  A.~Rockett, I.~Petrov, and J.~E. Greene, \emph{Physical properties of
  epitaxial {ZrN/MgO(001)} layers grown by reactive magnetron sputtering}, J.
  Vac. Sci. Technol. A \textbf{31}, 061516 (2013).
\bibAnnoteFile{Mei_jvsta_2013}

\bibitem{Atkins_prs_1966}
A.~G. Atkins and D.~Tabor, \emph{Hardness and deformation properties of solids
  at very high temperatures}, Proc. Roy. Soc. A \textbf{292}, 441--459 (1966).
\bibAnnoteFile{Atkins_prs_1966}

\bibitem{Atkins_jim_1966}
A.~G. Atkins, A.~Silverio, and D.~Tabor, J. Inst. Metals \textbf{94}, 369
  (1966).
\bibAnnoteFile{Atkins_jim_1966}

\bibitem{Chen_jac_2003}
C.~H. Chen, Y.~Xuan, and S.~Otani, \emph{Temperature and loading time
  dependence of hardness of {LaB$_6$, YB$_6$ and TiC} single crystals}, J.\
  Alloys\ Compd. \textbf{350}, L4 -- L6 (2003).
\bibAnnoteFile{Chen_jac_2003}

\bibitem{Ledbetter_jap_2006}
H.~Ledbetter and A.~Migliori, \emph{A general elastic-anisotropy measure}, J.\
  Appl.\ Phys. \textbf{100}, 063516 (2006).
\bibAnnoteFile{Ledbetter_jap_2006}

\bibitem{Zener_book}
C.~Zener, \emph{Elasticity and Aneslaticity of Metals} (University of Chicago,
  Chicago, 1948).
\bibAnnoteFile{Zener_book}

\bibitem{Ranganathan_prl_2008}
S.~I. Ranganathan and M.~Ostoja-Starzewski, \emph{Universal Elastic Anisotropy
  Index}, Phys.\ Rev.\ Lett. \textbf{101}, 055504 (2008).
\bibAnnoteFile{Ranganathan_prl_2008}

\bibitem{Kube_aipa_2016}
C.~C.~M.~Kube, \emph{Elastic anisotropy of crystals}, AIP Adv. \textbf{6},
  095209 (2016).
\bibAnnoteFile{Kube_aipa_2016}

\bibitem{Marmier_cpc_2010}
A.~Marmier, Z.~A.~D. Lethbridge, R.~I. Walton, C.~W. Smith, S.~C. Parker, and
  K.~E. Evans, \emph{{ElAM}: {A} computer program for the analysis and
  representation of anisotropic elastic properties}, Comput.\ Phys.\ Commun.
  \textbf{181}, 2102 -- 2115 (2010).
\bibAnnoteFile{Marmier_cpc_2010}

\bibitem{Ortiz_jcp_2013}
A.~U. Ortiz, , A.~Boutin, A.~H. Fuchs, and F.~X. Coudert, \emph{Metal–organic
  frameworks with wine-rack motif: What determines their flexibility and
  elastic properties?}, J.\ Chem.\ Phys. \textbf{138}, 174703 (2013).
\bibAnnoteFile{Ortiz_jcp_2013}

\bibitem{Gaillac_jpcm_2016}
R.~Gaillac, P.~Pullumbi, and F.~X. Coudert, \emph{{ELATE}: an open-source
  online application for analysis and visualization of elastic tensors}, J.\
  Phys.:\ Condens.\ Matter \textbf{28}, 275201 (2016).
\bibAnnoteFile{Gaillac_jpcm_2016}

\bibitem{ashcroft}
N.~W. Ashcroft and N.~D. Mermin, \emph{Solid State Physics} (Holt-Saunders,
  Philadelphia, 1976).
\bibAnnoteFile{ashcroft}

\bibitem{Grimvall_1999}
G.~Grimvall, \emph{Thermophysical Properties of Materials} (Elsevier, 1999).
\bibAnnoteFile{Grimvall_1999}

\bibitem{Jain_jac_2010}
A.~Jain, R.~Pankajavalli, S.~Anthonysamy, K.~Ananthasivan, R.~Babu, V.~Ganesan,
  and G.~S. Gupta, \emph{Determination of the thermodynamic stability of
  TiB$_2$}, J.\ Alloys\ Compd. \textbf{491}, 747 -- 752 (2010).
\bibAnnoteFile{Jain_jac_2010}

\bibitem{Chase_AIP_1998}
M.~W. {Chase,~Jr.}, \emph{{NIST}-{JANAF} Thermochemical Tables} (American
  Chemical Society and American Institute of Physics for the National Institute
  of Standards and Technology, Woodbury, NY, 1998), 4th edn.
\bibAnnoteFile{Chase_AIP_1998}

\bibitem{Zimmermann_jacers_2008}
J.~W. Zimmermann, G.~E. Hilmas, G.~W. Fahrenholtz, R.~B. Dinwiddie, W.~D.
  Porter, and H.~Wang, \emph{Thermophysical Properties of {ZrB2} and
  {ZrB$_2$–SiC} Ceramics}, J.\ Amer.\ Ceram.\ Soc. \textbf{91}, 1405--1411
  (2008).
\bibAnnoteFile{Zimmermann_jacers_2008}

\bibitem{Westrum_jct_1977}
E.~F. Westrum and G.~Feick, \emph{Heat capacities of {HfB$_{2.035}$ and
  HfC$_{0.968}$} from 5 to 350~{K}}, J. Chem. Thermodyn. \textbf{9}, 293 -- 299
  (1977).
\bibAnnoteFile{Westrum_jct_1977}

\bibitem{TCRC_book}
H.~L. Schick, ed., \emph{Thermodynamics of Certain Refractory Compounds}
  (Academic Press, New York, USA, 1966).
\bibAnnoteFile{TCRC_book}

\bibitem{UHTCs_book}
R.~Loehman, E.~Corral, H.~P. Dumm, P.~Kotula, and R.~Tandon, eds., \emph{Ultra
  High Temperature Ceramics for Hypersonic Vehicle Applications}
  (SAND2006‐2925, Albuquerque, USA, 2006).
\bibAnnoteFile{UHTCs_book}

\bibitem{Lengauer_jac_1995}
W.~Lengauer, S.~Binder, K.~Aigner, P.~Ettmayer, A.~Guillou, J.~Debuigne, and
  G.~Groboth, \emph{Solid state properties of group IVb carbonitrides}, J.\
  Alloys\ Compd. \textbf{217}, 137 -- 147 (1995).
\bibAnnoteFile{Lengauer_jac_1995}

\bibitem{Turchanin_sssr_1982}
A.~G. Turchanin and A.~E. Polyakov, \emph{Thermodynamic properties of hafnium
  carbide in 0-3000~{K} temperature interval}, Izv. Akad. Nauk SSSR, Neorg.
  Mater.; (USSR) \textbf{18:3}, 404--406 (1982).
\bibAnnoteFile{Turchanin_sssr_1982}

\bibitem{Storms1967refractory}
E.~K. Storms, \emph{The Refractory Carbides}, Refractory materials : a series
  of monographs (Academic Press, 1967).
\bibAnnoteFile{Storms1967refractory}

\bibitem{Westrum_jtac_2002}
E.~F. Westrum and J.~A. Sommers, \emph{Heat capacity of hafnium mononitride
  from temperatures of 5 to 350~{K}: An estimation procedure}, J. Therm. Anal.
  Calorim. \textbf{69}, 103 -- 112 (2002).
\bibAnnoteFile{Westrum_jtac_2002}

\bibitem{Ajami_jlcm_1974}
F.~I. Ajami and R.~K. MacCrone, \emph{Thermal expansion, Debye temperature and
  Gruneisen constant of carbides and nitrides}, J.\ Less-Common\ Met.
  \textbf{38}, 101 -- 110 (1974).
\bibAnnoteFile{Ajami_jlcm_1974}

\bibitem{BlancoGIBBS2004}
M.~A. Blanco, E.~Francisco, and V.~Lua{\~n}a, \emph{{GIBBS}:
  isothermal-isobaric thermodynamics of solids from energy curves using a
  quasi-harmonic Debye model}, Comput.\ Phys.\ Commun. \textbf{158}, 57--72
  (2004).
\bibAnnoteFile{BlancoGIBBS2004}

\bibitem{curtarolo:art96}
C.~Toher, J.~J. Plata, O.~Levy, M.~{de~Jong}, M.~D. Asta, M.~{Buongiorno
  Nardelli}, and S.~Curtarolo, \emph{High-throughput computational screening of
  thermal conductivity, {D}ebye temperature, and {G}r\"{u}neisen parameter
  using a quasiharmonic {D}ebye model}, Phys.\ Rev.\ B \textbf{90}, 174107
  (2014).
\bibAnnoteFile{curtarolo:art96}

\bibitem{AIP_Handbook}
D.~E. Gray and A.~A. Bennett, \emph{American Institute of Physics Handbook},
  McGraw-Hill handbooks (McGraw-Hill, 1972).
\bibAnnoteFile{AIP_Handbook}

\bibitem{Lawson_pm_2007}
A.~C. Lawson, D.~P. Butt, J.~W. Richardson, and J.~Li, \emph{Thermal expansion
  and atomic vibrations of zirconium carbide to 1600~{K}}, Philos. Mag.
  \textbf{87}, 2507--2519 (2007).
\bibAnnoteFile{Lawson_pm_2007}

\bibitem{Touloukian_Serie}
Y.~S. Touloukian, R.~W. Powell, C.~Y. Ho, and P.~G. Klemens,
  \emph{Thermophysical Properties of Matter - the TPRC Data Series.}
  (IFI/Plenum, 1970-1979).
\bibAnnoteFile{Touloukian_Serie}

\bibitem{Aigner_jac_1994}
K.~Aigner, W.~Lengauer, D.~Rafaja, and P.~Ettmayer, \emph{Lattice parameters
  and thermal expansion of {Ti(C$_x$N$_{1-x}$), Zr(C$_x$N$_{1-x}$),
  Hf(C$_x$N$_{1-x}$) and TiN$_{1-x}$ from 298 to 1473~K} as investigated by
  high-temperature X-ray diffraction}, J.\ Alloys\ Compd. \textbf{215}, 121 --
  126 (1994).
\bibAnnoteFile{Aigner_jac_1994}

\bibitem{Bogdanov_sov_1982}
V.~S. Bogdanov, V.~S. Neshpor, Y.~D. Kondrashev, A.~B. Goncharuk, and A.~N.
  Pityulin, Sov. Powder. Metall. Metal. Ceram. \textbf{21}, 412 (1982).
\bibAnnoteFile{Bogdanov_sov_1982}

\bibitem{Lonnberg_jlcm_1988}
B.~L\"onnberg, \emph{Thermal expansion studies on the group {IV–VII}
  transition metal diborides}, J.\ Less-Common\ Met. \textbf{141}, 145 -- 156
  (1988).
\bibAnnoteFile{Lonnberg_jlcm_1988}

\bibitem{Paxton_jc_2016}
W.~A. Paxton, T.~E. {\"O}zdemir, I.~{\c{S}}avkl{\i}y{\i}ld{\i}z, T.~Whalen,
  H.~Bi{\c{c}}er, E.~K. Akdo{\u{g}}an, Z.~Zhong, and T.~Tsakalakos,
  \emph{Anisotropic Thermal Expansion of Zirconium Diboride: An
  Energy-Dispersive {X}-Ray Diffraction Study}, J. Ceram. \textbf{2016},
  8346563 (2016).
\bibAnnoteFile{Paxton_jc_2016}

\end{thebibliography}
\end{document}